%% file: cocoonF13_ms_astro_ph.tex
\documentclass[twocolumn]{aastex63}
\usepackage{graphicx}
\usepackage{amsmath}
\usepackage{lineno}
\newcommand{\as}{^{\prime \prime}}
\newcommand{\kms}{\rm km~s^{-1}}
\newcommand{\bd}{\rm BD+46}
\newcommand{\ceo}{\rm C^{\rm 18}O}
\newcommand{\tco}{\rm ^{\rm 13}CO}
\newcommand{\mspc}{M_{\odot} ~\rm pc^{-1}}
\newcommand{\nht}{N_{\rm H_{2}}}
\newcommand{\mlin}{M_{\rm line}}

\shorttitle{The Formation of Filaments and Dense cores in the Cocoon Nebula}
\shortauthors{Chung et al.}

\begin{document}

\title{The Formation of Filaments and Dense cores in the Cocoon Nebula (IC~5146)}
%\title{The Formation of Filament-13 in the Cocoon Nebula via the Collision with Radiation Shock Front and the Fragmentation into Cores}

\author{Eun Jung Chung}
\affiliation{Department of Astronomy and Space Science, Chungnam National University, Daejeon, Republic of Korea}
\affiliation{Korea Astronomy and Space Science Institute, 776 Daedeokdae-ro, Yuseong-gu, Daejeon 34055, Republic of Korea}

\author{Chang Won Lee} \affiliation{Korea Astronomy and Space Science Institute, 776 Daedeokdae-ro, Yuseong-gu, Daejeon 34055, Republic of Korea}

\author{Shinyoung Kim} \affiliation{Korea Astronomy and Space Science Institute, 776 Daedeokdae-ro, Yuseong-gu, Daejeon 34055, Republic of Korea}

\author{Mario Tafalla} \affiliation{Observatorio Astron\'omico Nacional (IGN), Alfonso XII 3, E-28014 Madrid, Spain}

\author{Hyunju Yoo} \affiliation{Department of Astronomy and Space Science, Chungnam National University, Daejeon, Republic of Korea}

\author{Jungyeon Cho} \affiliation{Department of Astronomy and Space Science, Chungnam National University, Daejeon, Republic of Korea}

\author{Woojin Kwon} \affiliation{Department of Earth Science Education, Seoul National University, 1 Gwanak-ro, Gwanak-gu, Seoul 08826, Republic of Korea} \affiliation{SNU Astronomy Research Center, Seoul National University, 1 Gwanak-ro, Gwanak-gu, Seoul 08826, Republic of Korea}

\begin{abstract}
We present 850~$\mu$m linear polarization and $\ceo~(3-2)$ and $\tco~(3-2)$ molecular line observations toward the filaments (F13 and F13S) in the Cocoon Nebula (IC~5146) using the JCMT POL-2 and HARP instruments. F13 and F13S are found to be thermally supercritical with identified dense cores along their crests. Our findings include that the polarization fraction decreases in denser regions, indicating reduced dust grain alignment efficiency. The magnetic field vectors at core scales tend to be parallel to the filaments, but disturbed at the high density regions. Magnetic field strengths measured using the Davis-Chandrasekhar-Fermi method are 58$\pm$31 and 40$\pm$9~$\mu$G for F13 and F13S, respectively, and it reveals subcritical and sub-Alfv\'enic filaments, emphasizing the importance of magnetic fields in the Cocoon region. Sinusoidal $\ceo~(3-2)$ velocity and density distributions are observed along the filaments' skeletons, and their variations are mostly displaced by $\sim1/4 \times$wavelength of the sinusoid, indicating core formation occurred through the fragmentation of a gravitationally unstable filament, but with shorter core spacings than predicted. Large scale velocity fields of F13 and F13S, studied using $\tco~(3-2)$ data, present V-shape transverse velocity structure. We propose a scenario for the formation and evolution of F13 and F13S, along with the dense cores within them. A radiation shock front generated by a B-type star collided with a sheet-like cloud about 1.4~Myr ago, the filaments became thermally critical due to mass infall through self-gravity $\sim$1~Myr ago, and subsequently dense cores formed through gravitational fragmentation, accompanied by the disturbance of the magnetic field.
\end{abstract}


\section{Introduction}

Filamentary molecular clouds and dense cores are known as sites of ongoing star formation, and the formation mechanisms of filaments and dense cores have been investigated through simulations and observations over the last few decades. For the filament formation, simulation studies reveal that filaments emerge in gravitationally unstable infinite sheets and at the intersections of colliding turbulent sheets. Their formation and evolution are aided by the influence of the magnetic field, mechanical feedback, and radiation feedback from the surrounding clouds \citep[see Chapter 5 of][and references therein]{hacar2023}. \citet{abe2021} categorized filament formation mechanisms into five types : (i) a type G filament forms by self-gravitational fragmentation in a shock compressed sheet-like cloud, (ii) a type I filament forms at the intersections of two sheets, (iii) a type O filament is made by the curved magnetohydrodynamic shock wave, (iv) a type C filament forms by converging gas flow along the magnetic field lines in the post-shock layers, and (v) a type S filament is formed when a small clump is stretched by turbulent shear flows. This indicates that the filament formation mechanism is closely related to the environment as well as the energy budget of turbulence, magnetic field, and gravity of the molecular cloud where the filament is created. 

%=========== FIGURE:IC5146 Herschel 250 micron
\begin{figure*} \epsscale{1.17} \includegraphics[width=1\textwidth]{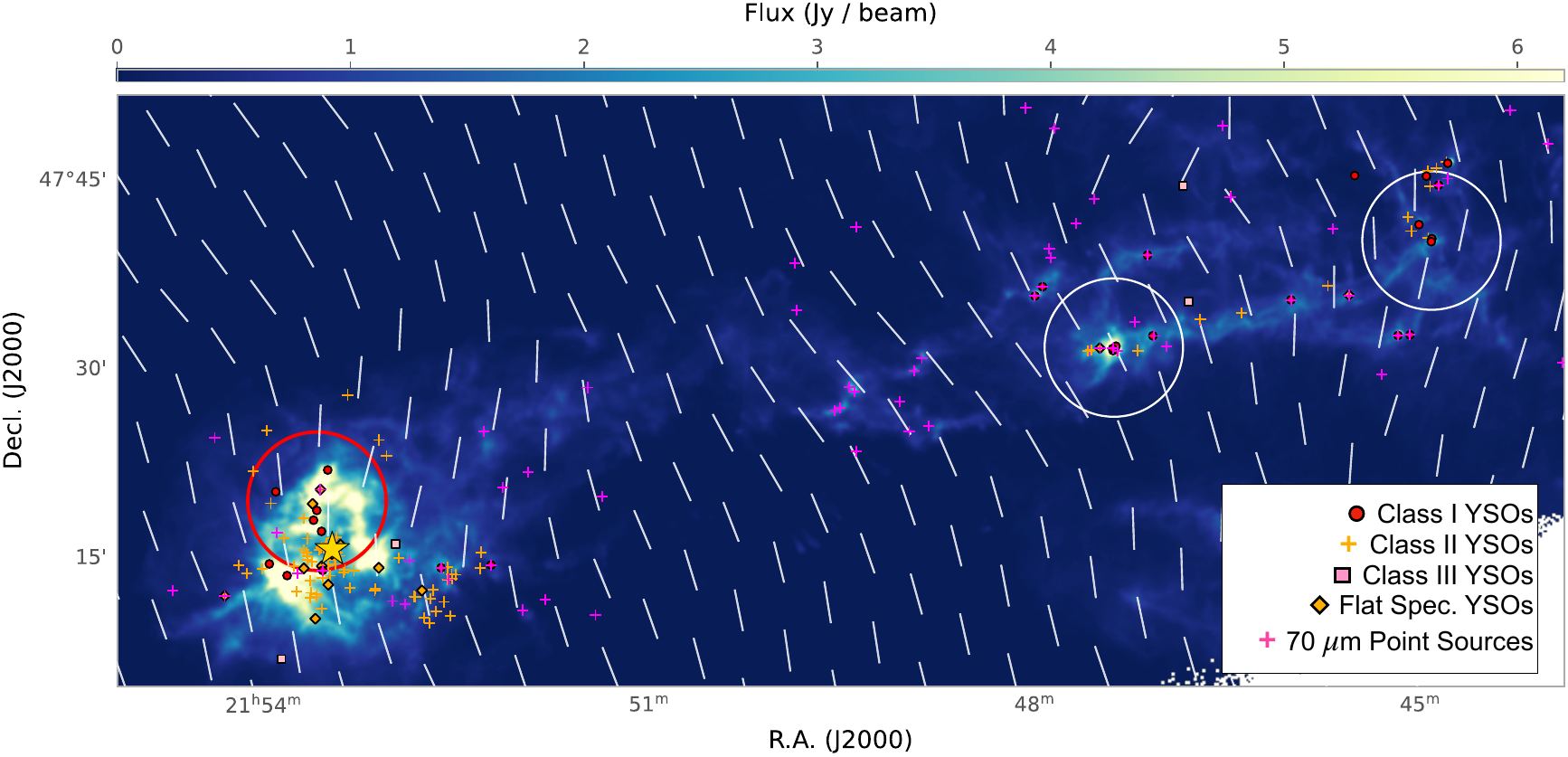} \caption{Herschel 250~$\mu$m image of IC~5146 \citep{andre2010}. The red circle on the east is the Filament-13 (F13) studied in this paper, and the white circles on the west indicate the eastern-hub (E-hub) and western-hub (W-hub) of the dark Streamer of IC~5146, which have been observed with JCMT/POL-2 \citep{wang2019,chung2022}. YSOs identified by Spitzer \citep{harvey2008} and 70~$\mu$m point sources by Herschel/PACS Point Source Catalogue \citep[HPPSC;][]{poglitsch2010} are presented. The big yellow star indicates the position of a B0 V star BD+46$^{\circ}~3474$ (BD+46). The white segments represent the magnetic field orientations inferred from the Planck 353~GHz polarization data \citep[Planck Collaboration Int][]{planck35}. \label{fig:obsregion}} \end{figure*}
%=========== 

Expanding {\sc Hi} shells, primarily formed by supernovae or {\sc Hii} regions, are suggested to provide sites for the formation of molecular filaments by observation and simulation studies \citep[e.g.,][]{dawson2008,bracco2020,hennebelle2008,inutsuka2015}. According to our current understanding, the large-scale shock compression resulting from multiple expanding shells can lead to the formation of flattened layers of molecular gas, with subsequent mass flows, driven by self-gravity or along magnetic fields, contributing to the creation of supercritical filamentary structures \citep{pineda2023}.

In this context, core formation can be modeled as the gravitational fragmentation of an isothermal equilibrium cylinder \citep[e.g.,][]{inutsuka1992}, although magnetic field and turbulence can also affect the formation of cores \citep[e.g.,][]{seifried2015,clarke2016,hanawa2017}. Hence, the observations toward filaments and dense cores have been performed to reveal the density and gas kinematic properties as well as the magnetic fields to examine their formation process \citep[e.g.,][]{arzoumanian2019,hacar2018,shimajiri2023,wang2022}.

IC~5146, also known as the Cocoon Nebula, is a nearby star-forming molecular cloud. The cloud exhibits a reflection nebula in the east, the Cocoon Nebula, and a dark cloud comprising multiple filaments in the west, referred to as the dark Streamer (Figure~\ref{fig:obsregion}). Due to their proximity in the plane of the sky, the Cocoon Nebula and the dark Streamer are generally studied together. However, the Cocoon Nebula and the dark Streamer have different star forming environments. Cocoon Nebula has $\sim$100 young stellar objects (YSOs), while the dark Streamer has only $\sim$20 YSOs \citep{harvey2008}. Also, a single B0 V star BD+46$^{\circ}~3474$ ~(hereafter \bd) resides at the center of Cocoon Nebula \citep[e.g.,][]{herbig2008,garcia2014}, surrounded by the {\sc Hii} region known as Sharpless 125 \citep[S125;][]{sharpless1959}. Based on optical and {\sc Hi} observations, it has been reported that BD+46 and the {\sc Hii} region are situated in front of the parent molecular cloud, suggesting that BD+46 likely formed first and created an ionized cavity, and the stars within the IC~5146 generated in a dense region of the molecular cloud located in the foreground, which subsequently dissipated after the emergence of BD+46 \citep[e.g.,][]{roger1982,herbig2002}.

The infrared dust continuum data reveal complex filamentary structures harboring most prestellar cores of IC~5146 \citep{arzoumanian2011}. \citet{chung2021} investigated the filaments and dense cores in IC~5146 with $\ceo~(1-0)$ and $\rm N_{2}H^{+}~(1-0)$ molecular line data, and found that most dense cores are located on the gravitationally supercritical filaments of IC~5146. Interestingly, the $\rm N_{2}H^{+}$ emission, known as one of the useful tracers of dense cores, is detected throughout the filaments of the dark Streamer, but it is merely detected in the filaments of the Cocoon Nebula. The filaments in the Cocoon Nebula have similar gravitational criticality to those of filaments in the dark Streamer, while they have slightly smaller nonthermal velocity dispersions normalized by the local sound speed than the filaments in the dark Streamer. Hence, \citet{chung2021} suggested that the thermal pressure and/or magnetic field support may be more significant in the filaments of the Cocoon Nebula than in those of the dark Streamer.

The magnetic field structures in IC~5146 have been studied by polarization observations at various wavelengths \citep[e.g., Planck Collaboration Int][]{planck35,wang2017,wang2019,chung2022}. The large scale magnetic fields appear to be nearly uniform and perpendicular to the elongated orientation of the IC~5146 Streamer \citep[e.g., Planck Collaboration Int][]{planck35}. However, on the core scale, the magnetic field orientations are revealed to be more complex. The white circles in Figure~\ref{fig:obsregion} shows the eastern-hub (E-hub) and western-hub (W-hub) of the dark Streamer where the magnetic fields have been investigated using the JCMT POL-2 polarimetry \citep{wang2019,chung2022}. The E-hub has a curved B-field indicating a possible drag by the gravitational contraction \citep{wang2019}. Meanwhile, the W-hub, which is likely more fragmented and evolved than the E-hub, shows much more complex B-field orientations \citep{chung2022}. The B-field strength of W-hub is 0.6$\pm $0.2~mG, and that of E-hub is 0.3$\pm$0.1~mG (recalculated in the same manner as that of W-hub by \citet{chung2022}). Although both hubs are magnetically supercritical, the E- and W-hubs are found to have slightly different fractions of gravitational ($E_{\rm G}$), turbulent ($E_{\rm K}$), and magnetic ($E_{\rm B}$) energy. This implies that the importance of magnetic field, turbulence, and gravity may change as filaments and dense cores evolve within different environments.

Recently, \citet{wang2020} used GAIA DR3 data to measure the distances to the Cocoon Nebula of 813$\pm$106~pc and to the dark Streamer of 600$\pm$100~pc. The distance of BD+46 is reported to be 800$\pm$80~pc \citep{garcia2014}, being well consistent to that of the Cocoon Nebula \citep{wang2020}. %The different star-forming properties and environment of the Cocoon Nebula compared to those of the dark Streamer give special importance to the Cocoon Nebula in the study of filament and core formation based on their evolutionary stages.

We carried out SCUBA-2/POL-2 and HARP observations toward the filamentary cloud regions in the Cocoon Nebula (the red circle area in Figure~\ref{fig:obsregion}) which consist of three filaments (named as F13, F13S, and F13W). These filaments are reported as thermally supercritical and sub/transonicsonic, and no $\rm N_{2}H^{+}$ emission is detected for these regions \citep{chung2021}. This filament system is thought to be a good laboratory to investigate the formation mechanism of filament and dense cores as having a different environment from that of the dark Streamer. 

The paper is organized as follows. We describe the observations and data reductions in Section~\ref{sec:obsdr}, and give the results of the observations in Section~\ref{sec:results}. We analyze and discuss in Sections~\ref{sec:anal} and \ref{sec:disc}, respectively. We summarize our results in Section~\ref{sec:summ}. \\

\section{Observations} \label{sec:obsdr}

\subsection{Polarization Observations}

Polarization observations were carried out with the JCMT POL-2 polarimeter between 2021 June 9 and 2022 August 9. The observations were performed toward F13 in the Cocoon Nebula using the standard SCUBA-2/POL-2 daisy mapping mode which %with a constant scanning speed of 4$\as~\rm s^{-1}$ and a data sampling rate of 8~Hz. 
covers the observing total area of a diameter $\sim 11^{\prime}$ with its central area of a diameter 3$^{\prime}$ having the best sensitivity. Its angular resolution is 14$\as$.1 at 850~$\mu$m wavelength (corresponding to $\sim$0.056~pc at a distance of 813~pc). The observations were made 21 times and the integration time of each observation was about 40 minutes resulting the total on source time of $\sim$14~hours under dry weather conditions with submillimeter opacity at 225~GHz ($\tau_{225 \rm GHz}$) ranging between 0.05 and 0.08. 

%=========== FIGURE:IC5146 Herschel 250 micron
\begin{figure*} %\epsscale{1.17}
\begin{center} \includegraphics[width=0.7\textwidth]{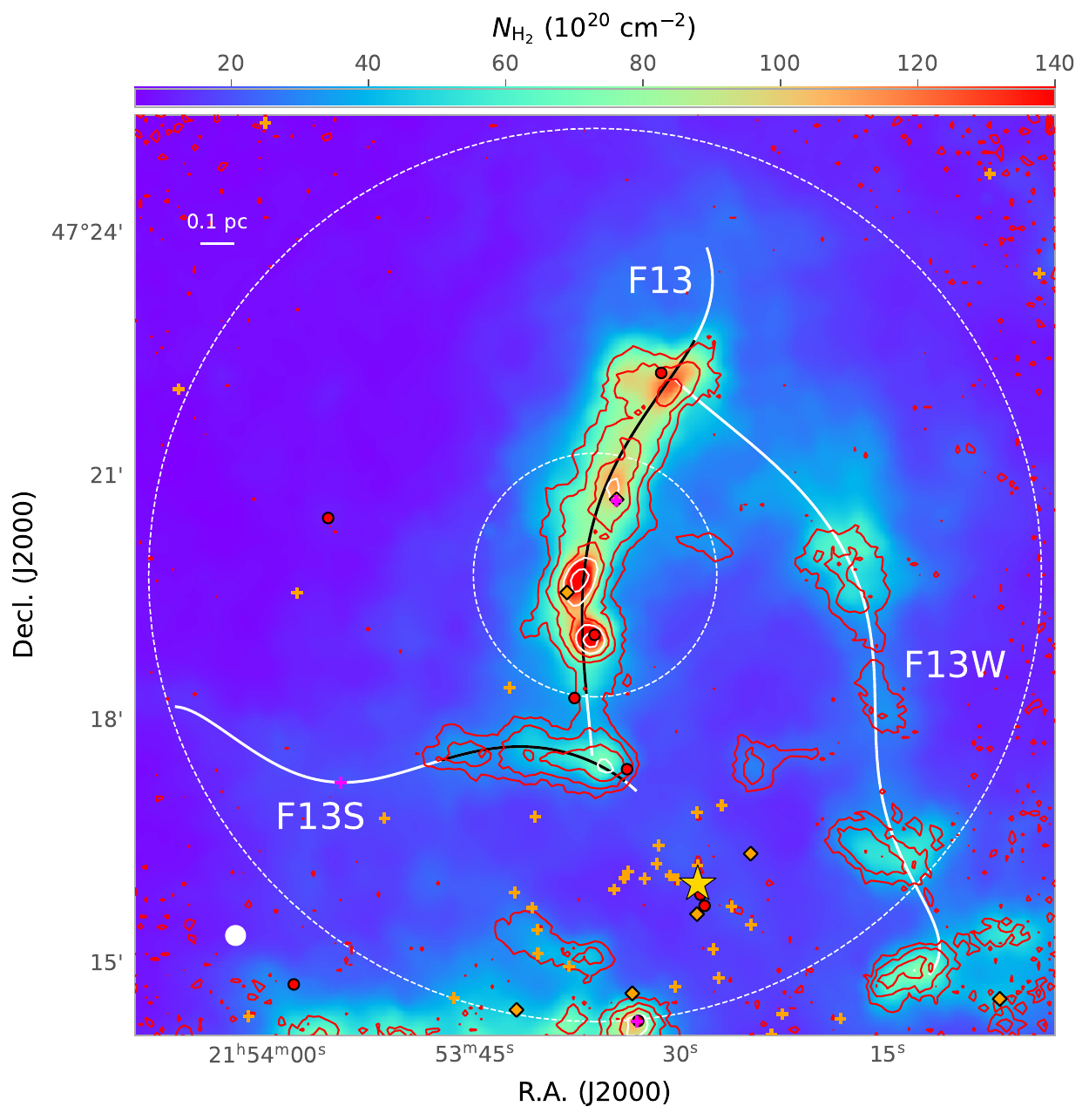} \end{center}
\caption{850~$\mu$m Stokes {\em I} map (red contours) on the H$_{2}$ column density map from the Herschel data \citep[color tone;][]{arzoumanian2011}. The contour levels of 850~$\mu$m emission are 3, 10, 20, 30, and 50$\times \sigma$ (1$\sigma$ = 4.0~mJy beam$^{-1}$). The yellow star at the south indicates the position of BD+46. The positions of YSOs identified by Spitzer \citep{harvey2008} and 70~$\mu$m point sources by Herschel/PACS Point Source Catalogue \citep[HPPSC;][]{poglitsch2010} are presented with the same as in Figure~\ref{fig:obsregion}. The dashed circles indicate the full area of POL-2 daisy map mode which covers a circular region of 11$^{\prime}$ diameter and the best sensitivity coverage of central 3$^{\prime}$ of the map, respectively. The white circle at the bottom left corner shows the POL-2 850~$\mu$m beam size of 14$\as$.1. The white curves present the crest of the filament identified from the Herschel H$_{2}$ column density data by \citet{arzoumanian2011}, which are named as F13, F13S, and F13W following \citet{chung2021}, and the black curves on them are the region where the data are analyzed in this paper.  \label{fig:obsresult}}
\end{figure*}
%=========== 

We used the $pol2map$ routine in the Sub-Millimetre User Reduction Facility ({\sc Smurf}) package of the Starlink software. Initially the raw bolometer time-streams are converted into {\em Q}, {\em U}, and {\em I} time-streams using the {\it calcqu} command and {\em I} time-streams of each observation are generated from the {\it makemap} command. In this process, a mask determined by the signal-to-noise ratio is used. Then, a pointing correction is applied and a mask determined from the co-added {\em I} map is used to generate an improved {\em I} map. Finally, {\em Q} and {\em U} maps are produced from their time-streams using the same mask in the previous step, and a final vector catalog is created. The final {\em I}, {\em Q}, and {\em U} maps have a bin-size of 4$\as$, and the vector catalog binned to 12$\as$ is used.

The `August 2019' IP model\footnote{\url{https://www.eaobservatory.org/jcmt/2019/08/new-ip-models-for-pol2-data/}} was applied for the instrumental polarization correction. A Flux Calibration Factor (FCF) of 668~Jy~pW$^{-1}$~beam$^{-1}$ is used for the 850 $\mu$m Stokes {\em I}, {\em Q}, and {\em U} data. The FCF is determined from the standard 850~$\mu$m SCUBA-2 flux conversion factor of 495~Jy~pW$^{-1}$~beam$^{-1}$ by multiplying a correction factor of 1.35 for the additional losses from POL-2 \citep{mairs2021}. The mean $rms$ level in Stokes {\em I} is 4.0~mJy~beam$^{-1}$, and the best $rms$ level within the central 3$^{\prime}$ region is 3.5~mJy~beam$^{-1}$. \\

%\subsection{$\ceo$ and $\tco~(3-2)$ Observations}
\subsection{Molecular Lines Observations}

$\ceo$ and $\tco~(3-2)$ molecular line observations toward F13 filaments by using the JCMT Heterodyne Array Receiver Programme \citep[HARP;][]{buckle2009} were performed to estimate their velocity centroids and dispersions. The main beam efficiency is 0.64 at 345~GHz \citep{buckle2009}. The data were taken using basket-weaved scan maps toward the central $3^{\prime} \times 7^{\prime}$ area between 2022 August 8 and September 1 in weather band~3 ($0.08 < \tau_{225 \rm GHZ} < 0.12$). The total observation time is 7.1~hours. The spatial and spectral resolutions for the $\ceo$ and $\tco~(3-2)$ observations are about 14$\as$ and 0.05~$\kms$, respectively. 

We used the `$\rm REDUCE\_SCIENCE\_NARROWLINE$' recipe of the ORAC-DR pipeline in the Starlink software. To increase the signal-to-noise ratio, we resampled the data cube to a channel width of 0.1~$\kms$ with a 1-d Gaussian kernel. The pixel size and the mean rms level of the final data cube are 7$\as$.3 and about 0.06~K$[T_{\rm A}^{\ast}]$, respectively.  \\

\section{Results} \label{sec:results}

%=========== FIGURE:long and radial column density profiles
\begin{figure*} \epsscale{1.17}
\includegraphics[width=0.407\textwidth]{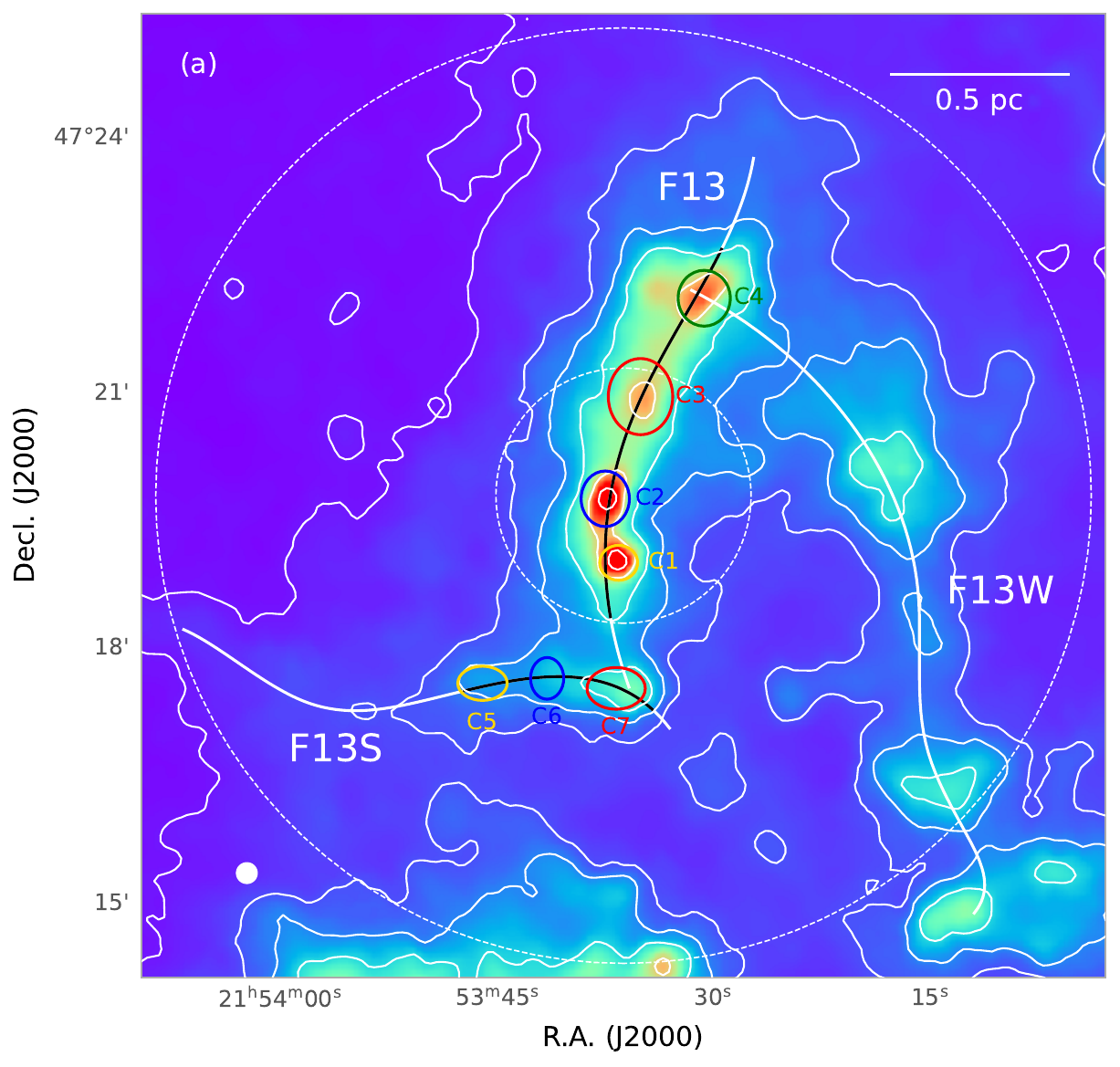}
\includegraphics[width=0.57\textwidth]{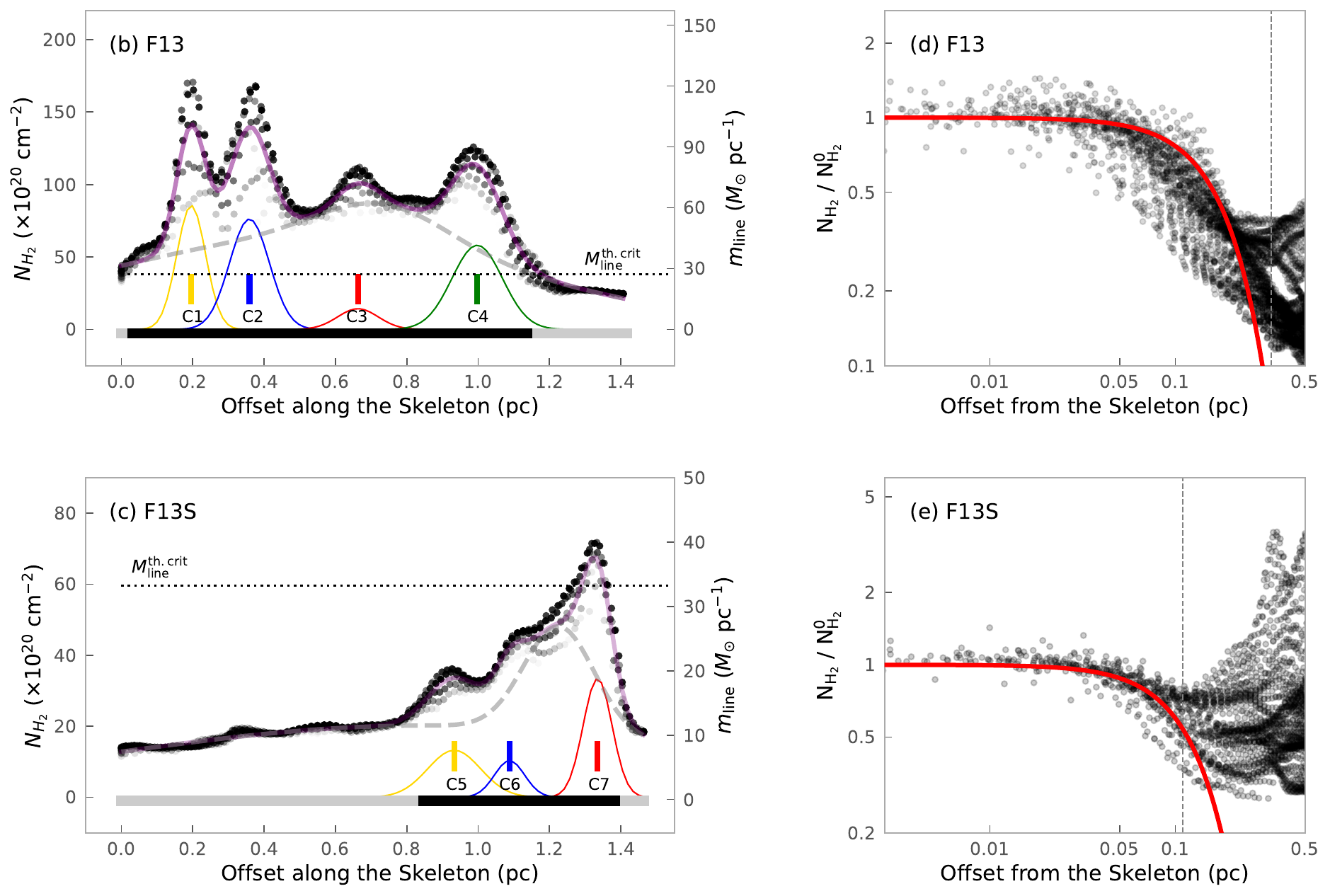}
\caption{Longitudinal and radial H$_{2}$ column density profiles of F13 and F13S. (a) Herschel H$_{2}$ column density map \citep{arzoumanian2019}. The color scale and curves are the same with that in Figure~\ref{fig:obsresult} and the contour levels are the H$_{2}$ column densities of 10, 20, 30, 50, 100, and 150$\times 10^{20}~\rm cm^{-2}$. The colored ellipses indicate the cores identified with {\sc FellWalker}. (b) and (c) Longitudinal H$_{2}$ column density profiles along the crest from south to north for F13 and from east to west for F13S, considering only pixels located at a distance $< 0.05$~pc from the crest. The colored curves represent the Gaussian models for the dense cores, while the thick dashed gray curve corresponds to the background filament material. The thick purple curve represents the sum of the Gaussian models and the filament material. The colored vertical segment indicates the peak position of the core's Gaussian model. On the right y-axis, the corresponding local line mass ($m_{\rm line}$) is presented, and the dotted horizontal lines indicate the thermal critical line mass ($m_{\rm line}^{\rm th.crit}$) at the mean temperature (please refer to the relevant text for more details). (d) and (e) Normalized radial column density profiles centered on the crest of filament and their Gaussian fits to estimate the filaments' widths. The Gaussian fits were performed only for the points in $r \leq r_{\rm th}$, where $r_{\rm th}$ is 0.33 and 0.11~pc for F13 and F13S, respectively. The red curve displays the Gaussian fit result and the vertical dashed line depicts the offset from the skeleton where $r=r_{\rm th}$. \label{fig:cdp}}
\end{figure*}
%===========

Figure~\ref{fig:obsresult} shows the Herschel H$_{2}$ column density map obtained from the Herschel Gould Belt survey\footnote{http://gouldbelt-herschel.cea.fr/archives} \citep[HGBS;][]{andre2010,arzoumanian2011} with 850~$\mu$m Stokes $I$ contours. The crests of filaments found from DisPerSE\footnote{http://www2.iap.fr/users/sousbie/web/html/indexd41d.html} are presented with solid curves \citep{arzoumanian2011}. They can be separated into three parts. We refer to the main filament as F13 following its discovery by \citet{chung2021}, while the remaining two as F13S and F13W based on their relative positions to F13. F13 exhibits the highest H$_{2}$ column density, followed by F13S and F13W, with a decreasing density. The 850~$\mu$m Stoke $I$ emissions show a good overall agreement with Herschel's H$_{2}$ column density distribution. Especially, the peak positions of 850~$\mu$m emission in F13 and F13S are well consistent to the H$_{2}$ column density peaks. \\

\subsection{Filaments and Dense Cores} \label{ssec:fandc}
 
We identified dense cores by applying the {\sc FellWalker} source extraction algorithm \citep{berry2015} to the 850~$\mu$m Stoke $I$ map. During the execution of this algorithm, we considered pixels with intensities higher than 0.5$\sigma$ to identify all dark cores. To classify an object as a genuine core, we set the minimum peak intensity threshold to 10$\sigma$ and the size threshold to 3 times the beam size of 14$\as$.1. We used 1.5$\sigma$ as the minimum dip threshold, which determines the neighboring peaks to be separated if the difference between the peak values and the minimum value (dip value) between the peaks is larger than the given threshold. From this, four and three cores were found in F13 and F13S, respectively. The cores in F13 are named as C1 to C4 from south to north, and those in F13S as C5 to C7 from east the west.

The identified cores are represented with ellipses over H$_{2}$ column density map provided by \citet{arzoumanian2011} in Figure~\ref{fig:cdp}(a) and their positions on the crest are marked with vertical lines over the longitudinal H$_{2}$ column density profile in Figure~\ref{fig:cdp}(b) and (c). As shown in the Figure, the 850~$\mu$m cores are well fitted with the $\nht$ peak positions. The mean projected separations of cores in F13 and F13S are 0.27$\pm 0.08$ and 0.20$\pm 0.05$~pc, respectively.

% TABLE : physical parameters of filaments 
\begin{deluxetable*}{lcccccccccc} 
\input{tbl_fils.tex}
\end{deluxetable*}
%===========

We estimated the physical quantities of F13 and F13S using the Herschel column density map. We measured the filament's length along the crest given by \citet{arzoumanian2011} within the enclosed region of $\nht > 50 \times 10^{20} \rm ~cm^{-2}$ for F13 and $\nht > 30 \times 10^{20} \rm ~cm^{-2}$ for F13S (depicted with black curves in Figure~\ref{fig:cdp}(a), (b), and (c)). The lengths of F13 and F13S are $1.13\pm0.09$ and $0.56\pm0.05$~pc, respectively. The length of F13 identified with the $\ceo~(1-0)$ molecular line data is 1.01~pc \citep{chung2021} which is comparable to that measured in this study.

As shown in Figure~\ref{fig:cdp}(b) and (c), the range of $\nht$ on the crest ($\nht^{0}$) of F13 is about 50 to 170$\times 10^{20}~\rm cm^{-2}$ and that of F13S is about 30 to 70$\times 10^{20}~\rm cm^{-2}$. The mean $\nht^{0}$ ($\bar N_{\rm H_{2}}^{0}$) of F13 and F13S are $\sim 93\pm27$ and $\sim 42\pm12 \times 10^{20}~\rm cm^{-2}$, respectively. 

To measure the filaments' widths ($W$), we used the normalized radial column density profiles ($\nht / \nht^{0}$) as presented in Figure~\ref{fig:cdp}(d) and (e). The profiles show a smooth decrement of column density with the increasing radial distance and some bumps at $r \gtrsim 0.2$~pc. The bumps are due to the substructure near C4 in the case of F13 and due to the other filament materials at the junctions of F13S and F13 as well as F13 and F13W in Figure~\ref{fig:cdp}(a). To avoid the effect of other filaments materials on the measurement, we set the threshold radius ($r_{\rm th}$) and performed Gaussian fitting only for the points at $r \leq r_{\rm th}$. We applied $r_{\rm th} = 0.33$ and 0.11~pc for F13 and F13S, respectively, at the point where $\nht$ begins to increase as the distance grows. The widths are estimated from the FWHM of the Gaussian ($W_{\rm uncorr}$) by correcting the beam smearing effect ($\sqrt{W_{\rm uncorr}^{2} - \theta_{\rm beam}^{2}}$ where $\theta_{\rm beam}=14\as.1$), and $W$ are $0.32\pm0.07$ and $0.23\pm0.03$~pc for F13 and F13S, respectively. These values are slightly larger than or similar to those obtained with the $\ceo~(1-0)$ molecular line data \citep[0.26~pc;][]{chung2021}.

The mass per unit length ($\mlin = M / L$) is commonly examined as an indicator of filament instability, similar to the Jeans mass for spherical systems. On the right y-axis of Figure~\ref{fig:cdp}(b) and (c), we presented the corresponding local line mass ($m_{\rm line}$) estimated from the H$_{2}$ column density by multiplying the filament width ($N_{\rm H_{2}}^{0} \times W$). The line mass is found to range between $\sim 10$ and $120 ~\mspc$. In a hydrostatic isothermal cylinder model, the critical line mass ($\mlin^{\rm th.crit}$) represents the equilibrium point where the thermal pressure balances with the gravitational collapse. It can be calculated using the equation:
\begin{equation}
	M_{\rm line}^{\rm th.crit} = \frac{2 c_{\rm s}^{2}}{G}, \label{eq:mline}
\end{equation}
where $c_{\rm s}$ is the sound speed \citep{ostriker1964}. At the mean dust temperatures of 19.8$\pm$1.2 and 24.2$\pm$1.0~K of F13 and F13S regions, $\mlin^{\rm th.crit}$ of F13 and F13S are $\sim$27$\pm$2 and 33$\pm$1 $\mspc$, respectively. The local line mass of F13 is found to be larger than the critical line mass along the entire crest, whereas that of F13S is higher than the critical line mass only near the westernmost core (C7). 

We measured H$_{2}$ number density ($n_{\rm H_{2}}$) of each filament by assuming the cylindrical geometry of filament having the diameter of estimated width. $n_{\rm H_{2}}$ at the distance, $r$, from the skeleton is calculated with the following equation:
\begin{equation}
	n_{\rm H_{2}} = \frac{N_{\rm H_{2}}}{l_{\rm los}}, \label{eq:ndh2}
\end{equation}
where the line-of-sight length, $l_{\rm los}$, is
\begin{equation}
	l_{\rm los} = 2 \times \sqrt{ \left( \frac{W}{2} \right)^{2} - r^{2} }, 
\end{equation}
where $W$ is the filament's width. The estimated H$_{2}$ number densities along the skeleton of F13 and F13S range 8$-17\times 10^{3}~\rm cm^{-3}$ and 6$-8\times 10^{3}~\rm cm^{-3}$, respectively. The mean values ($\bar n_{\rm H_{2}}^{0}$) are $11\pm2\times 10^{3}~\rm cm^{-3}$ and $7\pm1 \times 10^{3}~\rm cm^{-3}$ for F13 and F13S, respectively. The calculated physical parameters of filaments are listed in Table~\ref{tbl:fila}. \\

\subsection{Polarization Properties} \label{ssec:polp}

%=========== FIGURE: Pol vectors
\begin{figure*} \begin{center} %\epsscale{1.17}
\includegraphics[width=0.85\textwidth]{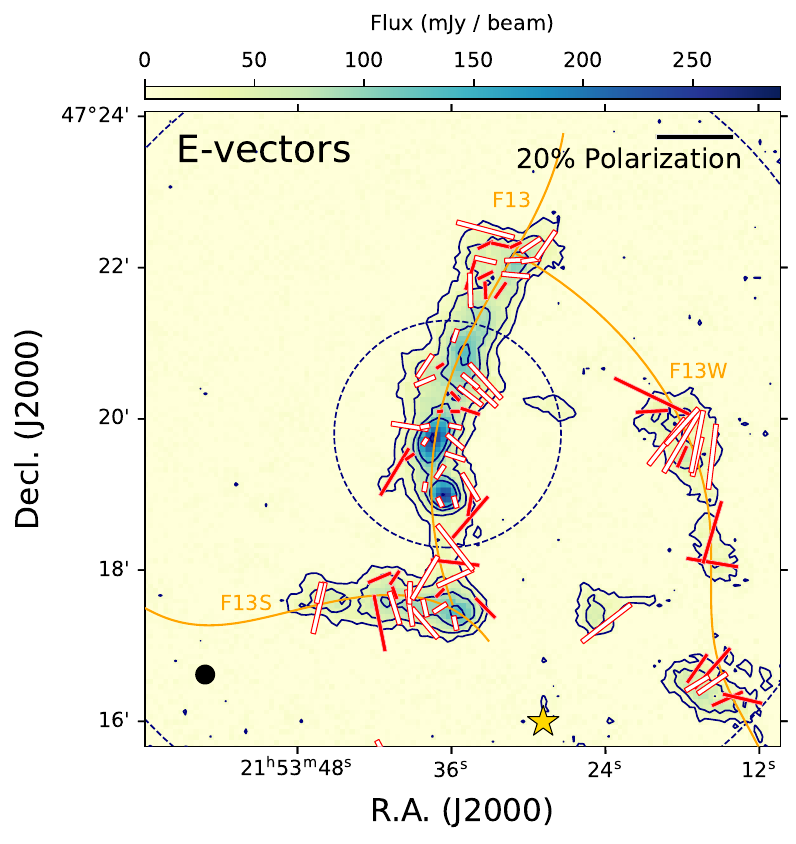} \end{center}
\caption{The polarization segments on the 850~$\mu$m Stokes $I$ intensity map (color scale and contours). The segments are selected with the criteria $\sigma_{I}/I \geq 10$ and $\sigma_{P}/P \geq 2$. The white and red bars are for the polarization vectors with $2 \leq P/ \sigma_{P} < 3$ and $P/\sigma_{P} \geq 3$, respectively. A 20~\% line segment is presented for reference on the top right corner. The black circle at the bottom left corner is the POL-2 850~$\mu$m beam size of 14.1$\as$. The orange curves are the skeletons of filaments and the yellow star indicates the position of BD+46. \label{fig:polvectors}}
\end{figure*}
%=========== 

Using the obtained $Q$ and $U$ maps, polarization intensity ($PI$), polarization fraction ($P$), and polarization angle ($PA$) were obtained. $PI$ is the quadratic sum of $Q$ and $U$ ($PI = \sqrt{Q^{2} + U^{2}}$), and the noises of $Q$ and $U$ always make a positive bias in $PI$ and $P$ \citep{vaillancourt2006}. Hence, a correction was performed with the modified asymptotic estimator \citep{plaszczynski2014} and the debiased polarization intensity was estimated with the following equation:
\begin{equation}
	PI = \sqrt{Q^{2} + U^{2}} - \sigma^{2} \frac{1 - e^{-(Q^{2} + U^{2})/\sigma^{2}}}{2\sqrt{Q^{2} + U^{2}}}.   \label{eq:pi}
\end{equation}
\noindent The noise bias $\sigma$ is calculated from %weighted mean of the variances on {\em Q} and {\em U}, $\sigma^{2}$, was calculated with 
the following equation \citep{plaszczynski2014}:
\begin{equation}
	\sigma^{2} = \frac{Q^{2} \sigma_{U}^{2} + U^{2} \sigma_{Q}^{2}}{Q^{2} + U^{2}},
\end{equation}
where $\sigma_{Q}$ and $\sigma_{U}$ are the standard errors in {\em Q} and {\em U}, respectively.

%=========== FIGURE: i - pi relation
\begin{figure*} %\epsscale{1.17}
\includegraphics[width=1\textwidth]{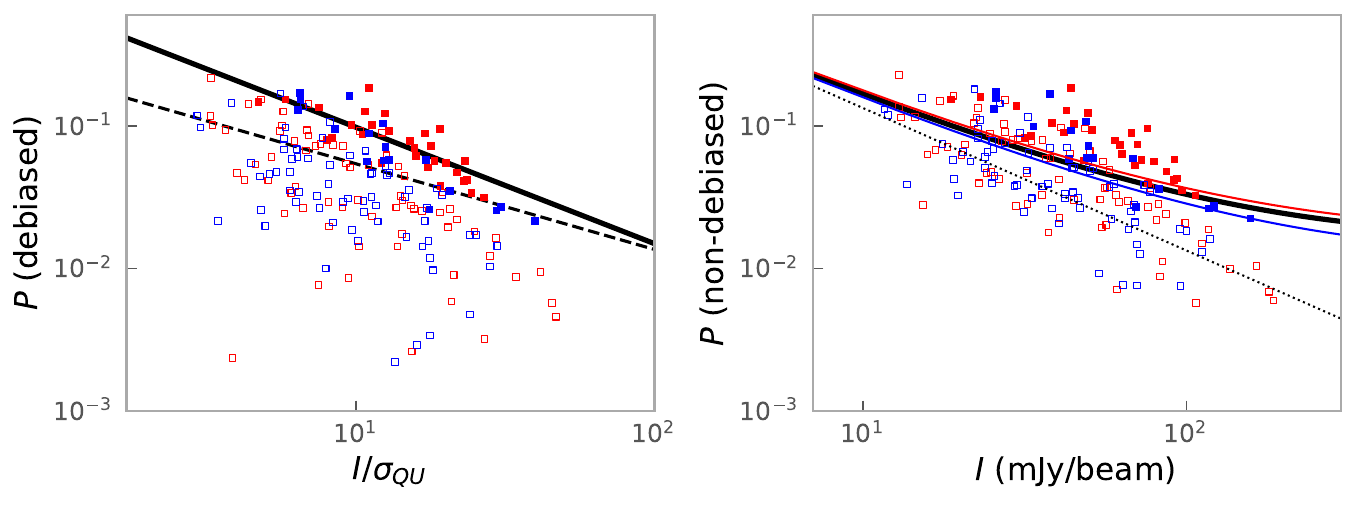}
\caption{Relationship between the polarization fraction and the Stokes $I$ intensity. The open and filled squares denote the polarization segments with $P/\sigma_{P} < 3$ and $P/\sigma_{P} \geq 3$, respectively. The red and blue squares indicate the vectors of closer and far regions, respectively (see Section~\ref{ssec:pi}). {\it Left:} the debiased polarization fraction as a function of the normalized Stoke $I$ intensity by $\sigma_{QU}$ (the mean of $\sigma_{Q}$ and $\sigma_{U}$). The dashed and solid lines are the best fit to a single power-law function, $P(\rm debiased) = \it P_{\sigma_{QU}}(I/\sigma_{QU})^{-\alpha}$, with polarization segments of $I/\sigma_{I} \geq 10$ and those of $I/\sigma_{I} \geq 10$ and $P/\sigma_{P} \geq 3$, respectively. {\it Right:} Relationship between the non-debiased polarization fraction and the Stokes $I$ intensity. The solid black line is the best-fit Ricean-mean model with all the polarization segments, and the dotted line is the null hypothesis case ($\alpha=0$). The red and blue curves are the best fit Ricean-mean model with the vectors of closer and far regions from the BD+46, respectively. \label{fig:ipf}}
\end{figure*}
%=========== 

The debiased polarization fraction $P$ was measured by 
\begin{equation}
	P = PI/I   \label{eq:p},	
\end{equation}
 and its uncertainty was calculated by propagating the standard errors of $PI$ ($\sigma$) and $I$ ($\sigma_{I}$) with the equation of \begin{equation}
	\sigma_{P} \simeq \sqrt{\frac{\sigma^{2}}{I^{2}} + \frac{\sigma_{I}^{2}(Q^{2} + U^{2})}{I^{4}}}.
\end{equation}

The polarization position angle, $\theta$, can be calculated by 
\begin{equation}
	\theta = \frac{1}{2} \rm arctan \frac{\it U}{\it Q}, \label{eq:theta}
\end{equation}
\noindent from the definition of $Q$ and $U$ parameters of 
\begin{equation}
	Q = I \times P \times \rm cos(2\theta) \label{eq:q}
\end{equation}
\noindent and
\begin{equation}
	U = I \times P \times \rm sin(2\theta) .\label{eq:q}
\end{equation}
\noindent The uncertainty of $\theta$ was calculated by the equation \citep[e.g.,][]{ngoc2021}:
\begin{equation}
	\sigma_{\theta} = \frac{1}{2} \sqrt{\frac{Q^{2} \sigma_{U}^{2} + U^{2} \sigma_{Q}^{2}}{(Q^{2} + U^{2})^{2}}}.
\end{equation}

%===========================
\begin{deluxetable*}{lccccc} 	
\input{tbl_alpha.tex}
\end{deluxetable*}
%=================

Figure~\ref{fig:polvectors} shows the polarization vectors on the Stokes $I$ image. With the selection criteria of $I/\sigma_{I} \geq 10$ and $P / \sigma_{P} \geq 2$, 94 % 79 in the given figure
vectors are obtained (51 vectors having $P / \sigma_{P} \geq 3$). One noticeable feature is that the polarization vectors are mostly perpendicular to the skeletons of filaments, especially in F13S and F13W. The polarization fractions appear to be lower at the brighter region.

It is well known that the polarization fraction tends to decrease as the intensity increases with a power-law relation of $P \propto I^{-\alpha}$ \citep[e.g.,][]{pattle2019,whittet2008}. Dust polarization is observed likely because of the non-spherical shape of dust grains and the alignment of their minor axes parallel to the local magnetic field. The polarization fraction is related to the alignment efficiency of the dust grains which may be determined by the size and composition of dust. It is quite complex to infer the dust alignment efficiency from the observed polarization fraction because it can be affected by the dust opacity and the mixing of various magnetic fields with different strength and geometry along the line-of-sight direction. However, it still gives important information on the dust alignment efficiency, and the power-law index $\alpha$ is useful to demonstrate the polarization properties of a cloud. $\alpha=0$ implies that the dust in a cloud has the same alignment efficiency at all optical depths. $\alpha=0.5$ means that the efficiency of dust alignment linearly decreases as the optical depth increases. $\alpha=1$ is observed if the dust grains align only at the thin surface layer of the cloud but no dust grains align at higher densities \citep[e.g.,][]{whittet2008}. 

%=========== FIGURE: harp result
\begin{figure*} \epsscale{1.17}
\includegraphics[width=1\textwidth]{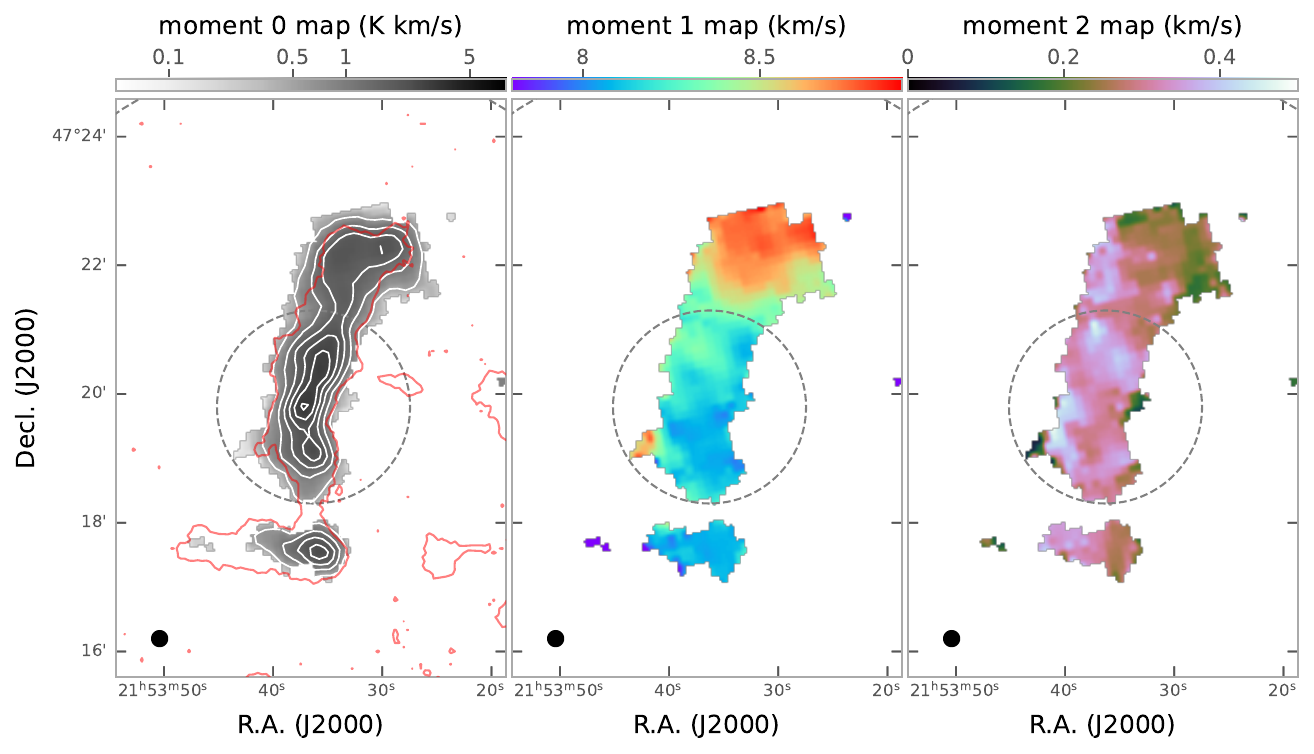}
\caption{$\ceo~(3-2)$ moment maps obtained from the HARP instrument of JCMT. In the moment 0 map, the integrated intensity contours of $\ceo$ are drawn with white color and the contour levels are 10, 20, 30, $\cdots$, 60$\times \sigma$ ($\sigma=0.053~\rm K~km~s^{-1}$). The red contour is the 3$\sigma$ level of the 850~$\mu$m Stokes $I$ intensity. The spatial resolution is 14$\as$, and the beam is given at the bottom left panel of the maps. %The dashed circles are the 3$^{\prime}$ central region.  
\label{fig:c18o}}
\end{figure*}
%=========== 

We presented the relationship between the intensity and polarization fraction in Figure~\ref{fig:ipf}. The left panel shows the debiased polarization fraction ($P_{\rm db}$) as a function of the normalized $I$ intensity divided by $\sigma_{QU}$ (the mean of $\sigma_{Q}$ and $\sigma_{U}$). We performed a least squares single power-law fit of $P_{\rm db} = P_{\sigma_{QU}} (I/\sigma_{QU})^{-\alpha}$, where $\sigma_{QU}$ is the representative rms noise in Stokes $Q$ and $U$ (and the reference intensity), and $P_{\sigma_{QU}}$ is the polarization fraction at the reference intensity ($\sigma_{QU}$). The fit results with vectors of $I/\sigma_{I} \geq 10$ and with those of $I/\sigma_{I} \geq 10$ and $P/\sigma_{P} \geq 3$ are overlaid with dashed and solid lines, respectively. The power-law index $\alpha$ of the fit using the polarization segments of $I/\sigma_{I} \geq 10$ and $P/\sigma_{P} \geq 3$ is $0.82\pm0.09$. 

%=========== FIGURE: longitudinal velocity field
\begin{figure*} %\epsscale{1.17}
\includegraphics[width=0.99\textwidth]{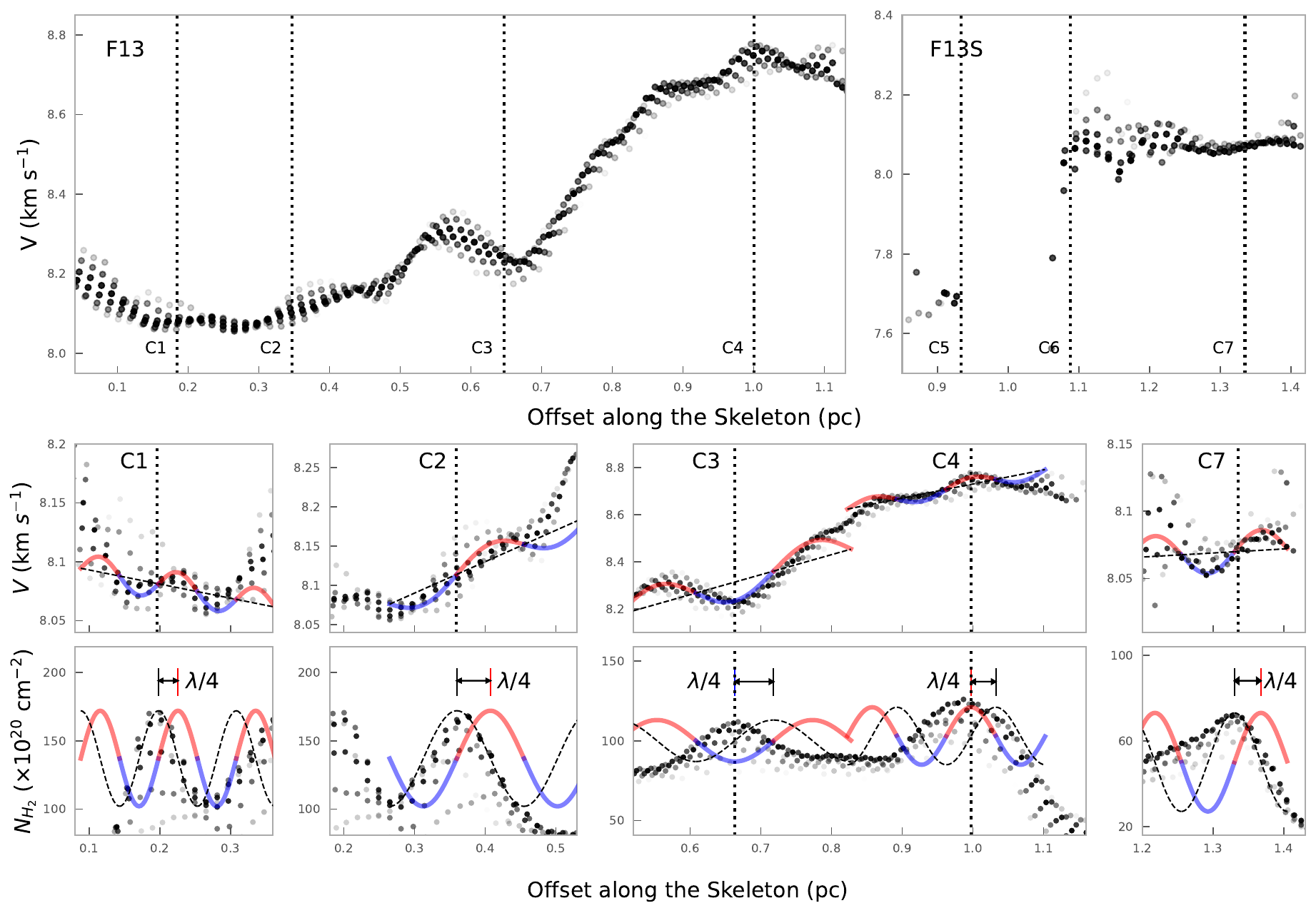}
\caption{Longitudinal $\ceo~(3-2)$ velocity (top and middle panels) and $N_{\rm H_{2}}$ variations (bottom panels) in F13 and F13S. The cores' positions are presented with dotted vertical lines. {\it Top :} The longitudinal velocity variations along the crest, from south to north for F13 and from east to west for F13S. Only pixels located within a distance $< 0.05$pc from the crest are considered. %The cores' positions on the crest are depicted with colored vertical lines. 
{\it Middle :} Same as in the top panels, but shown around each core. The curves depicted with red and blue are the best fit of the sinusoid to the radial velocity field and the dashed lines are the linear velocity gradients of the sinusoidal fits. {\it Bottom :} The longitudinal H$_{2}$ column density variations as shown in Figure~\ref{fig:cdp}, but around each core. The colored curve indicates the sinusoidal fit result in the middle panel, and its $\lambda/4$ shift is presented with the black dashed curve. \label{fig:longiv}}
\end{figure*}
%=========== 

We also obtained $\alpha$ using the Ricean-mean model \citep{pattle2019}. The single power-law model is appropriate for the data with high signal-to-noise ratio, but it is not valid for the low S/N data where the observed polarization fraction follows a Rice distribution and thus $\alpha$ may be overestimated \citep{pattle2019}. The Ricean-mean model was adopted to show that $\alpha$ can be well recovered by this model. We applied the Ricean-mean model to the non-debiased data with $I/\sigma_{I} \geq 10$ following the equation \citep{pattle2019}:
\begin{equation}
	P = \sqrt{\frac{\pi}{2}} \left( \frac{I}{\sigma_{QU}} \right)^{-1} \mathcal{L}_{\frac{1}{2}}\left[ -\frac{P_{\sigma_{QU}}^{2}}{2} \left( \frac{I}{\sigma_{QU}} \right)^{2(1-\alpha)} \right],
\end{equation}
where $\mathcal{L}_{\frac{1}{2}}$ is a Laguerre polynomial of order 1/2. 

The right panel of Figure~\ref{fig:ipf} shows the relationship between the non-debiased polarization fraction and the Stokes $I$ intensity with the best-fit Ricean-mean model. The obtained $\alpha$ and $P_{\sigma_{QU}}$ are $0.72\pm0.07$ and $0.33\pm0.08$, respectively. The fitting results are tabulated in Table~\ref{tbl:alpha}.

The expected $\alpha$ for the molecular clouds ranges between 0.5 and 1. The molecular clouds studied in the BISTRO survey have $\alpha$ of 0.8$-$1.0 with a single power-law model. $\alpha$ obtained from the Ricean-mean model of the BISTRO objects ranges between 0.3 and 0.7 \citep[e.g.,][]{pattle2019,wang2019,lyo2021,arzoumanian2021,ching2022,hwang2022}. 

The obtained $\alpha$ of the Cocoon Nebula from a single power-law function as well as from the Ricean-mean model lie well between the expected range of molecular cloud. $\alpha$ from a single power-law function for the vectors having $I/\sigma_{I} \geq 10$ and $P/\sigma_{P} \geq 3$ is quite close to that of the early B-type star LkH$\alpha$~101 region. %, but $\alpha$ from a single power-law function for the vectors having $I/\sigma_{I} \geq 10$ is quite lower than those of BISTRO targets. 
On the contrary, $\alpha$ from the Ricean-mean model falls on the higher end of the BISTRO's range. We discuss the relationship between $I$ and $P$ in Section~\ref{ssec:pi}. \\

\subsection{Velocity Field from the $\ceo$ and $\tco~(3-2)$ Data} \label{ssec:c18o}

Figure~\ref{fig:c18o} presents the $\ceo~(3-2)$ moment maps. The integrated intensity map shows that the distribution of $\ceo$ emission is well matched to that of 850~$\mu$m emission in F13, but in F13S region it is detected only at the dense western part. 

%=========== FIGURE: radial velocity field
\begin{figure*} %\epsscale{1.17}
\includegraphics[width=1\textwidth]{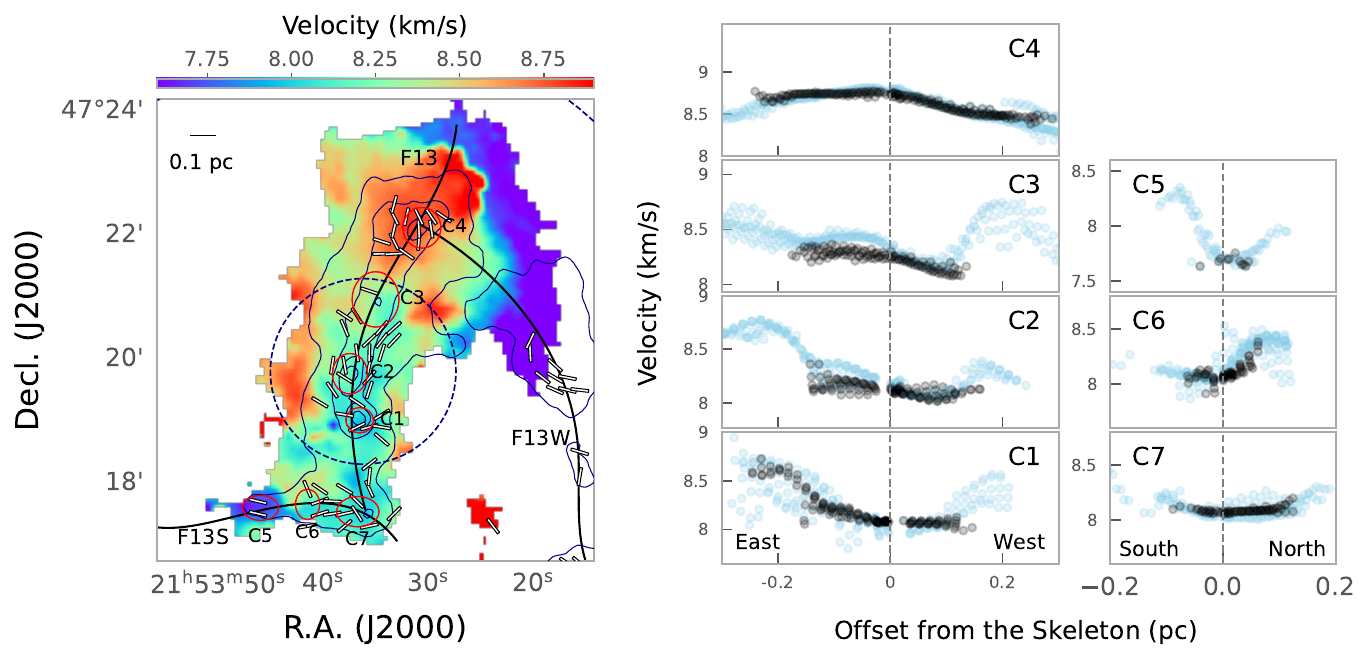}
\caption{Velocity field map traced by the $\tco~(3-2)$ molecular line data (left panel) and the radial velocity profiles centered on the cores are presented (right panels). {\it Left :} The $\tco(3-2)$ central velocity map derived by fitting the Gaussian function to the $\tco $ line profiles is displayed. Contour levels of H$_{2}$ column density are presented at 30, 70, 110, and 150$\times 10^{20}~\rm cm^{-2}$. The white segments represent the magnetic field vectors and the navy contours represent the Herschel column density. %, and the dashed circles are the same as in Figure~\ref{fig:bvectors}. 
{\it Right :} Presented are the transverse velocity profiles at each core's position from east to west for F13 and from south to north for F13S. The black and sky-blue circles indicate the velocities traced from the $\ceo$ and $\tco~(3-2)$ molecular lines, respectively.   \label{fig:vfields}}
\end{figure*}
%=========== 

The moment 1 map reveals a gradual increase in velocity, ranging from approximately 7.7~$\kms$ in the southeast, 8.0~$\kms$ in the south, to 8.8~$\kms$ in the north. The mean velocity gradient of F13 is $\sim 0.9~\kms ~\rm pc^{-1}$ and that of F13S is $\sim 1.0~\kms ~\rm pc^{-1}$. The velocity dispersions are in a range between $\sim$0.2 and 0.6~$\kms$ with the mean value of 0.30$\pm0.06~\kms$. We inspected the spectra and discovered that all of them exhibit a single velocity component.

Figure~\ref{fig:longiv} presents the $\ceo~(3-2)$ velocity structure along the skeleton of F13 and F13S. F13 exhibits a globally increasing velocity field from south to north, but with small oscillatory behaviors. The dotted vertical lines depict the cores' positions, and the velocity decreases and then increases around C3 up to C4 position along the skeleton from the south to north. In contrast, around the other cores, such abrupt changes in velocity direction are not observed.

A sinusoidal velocity distribution along the filament's skeleton has been reported in previous molecular line studies \citep[e.g.,][]{hacar2011,chung2019,chung2021,kim2022,shimajiri2023}. It is explained as the result of core formation via fragmentation of the filament, and the oscillatory fluctuations of velocity and density are expected to present a shift of $\lambda/4$, where $\lambda$ is the wavelength of both the density and velocity fluctuations \citep{hacar2011,kim2022,shimajiri2023}. 

The middle and bottom panels in Figure~\ref{fig:longiv} show the detailed longitudinal velocity and H$_{2}$ column density profiles around the cores. We fitted a sinusoid function to the velocity profile and overlaid it with the red and blue curves. The $\lambda/4$-shifted curve is depicted with a dashed line on the $\nht$ profile in the bottom panel. We found the best fit sinusoid for C1, C2, and C7, where the position of the density peak is shifted by $\lambda/4$ from the velocity peak position. The observed velocity structures around the three cores agree to those of core forming mass flow by filament fragmentation.

%=========== FIGURE: nonthermal velocity dispersion
\begin{figure*} %\epsscale{1.17}
\includegraphics[width=1\textwidth]{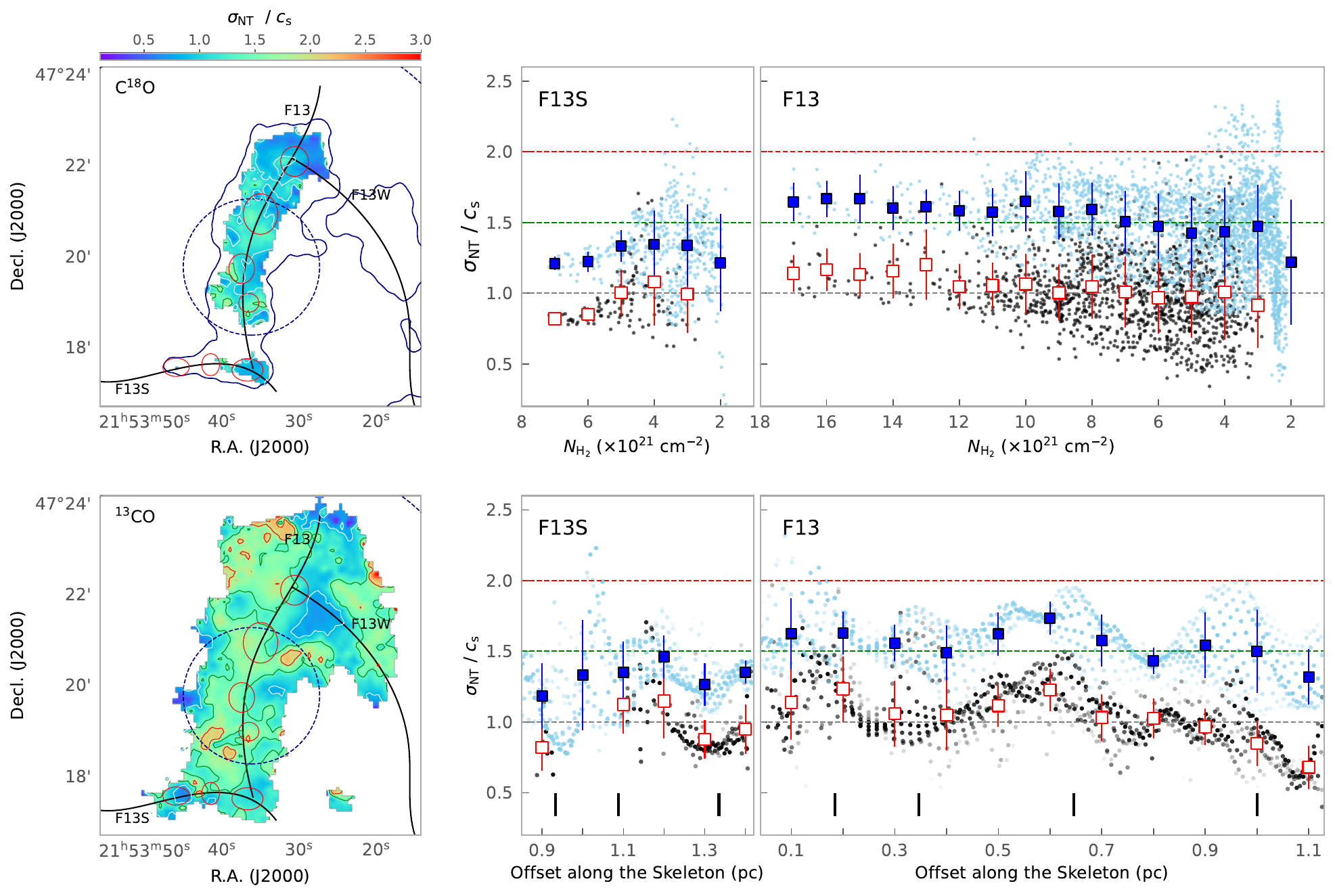}
\caption{The distribution of nonthermal velocity dispersion normalized with the sound speed ($\sigma_{\rm NT}/c_{\rm s}$). %$\sigma_{\rm NT}$ is estimated with the observed velocity dispersion of $\ceo$ and $\tco$ using Equation~\ref{eq:sigmaNT} and the sound speed is measured with the Herschel dust temperature map. 
{\it Left:} The maps of $\sigma_{\rm NT}/c_{\rm s}$ from the $\ceo$ (top) and $\tco$ line profiles (bottom). The white, green, and red contours are $\sigma_{\rm NT}/c_{\rm s} = 1$, 1.5, and 2, respectively. The navy contour on the $\sigma_{\rm NT}/c_{\rm s}$ map from the $\ceo$ is the 30$\times 10^{20}~\rm cm^{-2}$ level of Herschel H$_{2}$ column density. {\it Right:} $\sigma_{\rm NT}/c_{\rm s}$ as a function of H$_{2}$ column density (top) and the distance along the skeleton from east to west for F13S and from south to north for F13 (bottom). $\sigma_{\rm NT}/c_{\rm s}$ from $\ceo$ and $\tco$ are illustrated with black and skyblue colors, respectively, and their averaged values are presented with white and blue squares. The horizontal dashed lines indicate $\sigma_{\rm NT}/ c_{\rm s}$ equals to 1, 1.5, and 2. The positions of dense cores are denoted with balck vertical lines in the bottom panels. \label{fig:sigNT}}
\end{figure*}
%=========== 

The sinusoidal variation for other four cores was not clearly identified, as there were not enough $\ceo$ velocity data points for C5 and C6. C3 and C4 have the density peak position at the velocity minimum or maximum position. C3 and C4 have lower H$_{2}$ column densities than C1 and C2, but the line masses $M_{\rm line}$ near C3 and C4 are larger than the critical line mass (see Figure~\ref{fig:cdp} and Table~\ref{tbl:fila}). Hence, we can assume that the observed velocity fields around these two cores are also due to the mass inflow into cores along the filament, and we can infer the three-dimensional structure of the filament. Specifically, to have minimum or maximum velocity at the density peak, F13 must possess a bend along the line of sight, where the high-velocity regions are located closer and the low-velocity regions are situated farther away from us. 

Figure~\ref{fig:vfields} shows the velocity fields of F13 and F13S using $\tco~(3-2)$ data, which were simultaneously obtained with the HARP observations of $\ceo~(3-2)$ line. In a previous molecular line study toward this region, we found that the $\tco~(1-0)$ emission is well matched with the Herschel 250~$\mu$m emission, while the $\ceo~(1-0)$ emission is distributed only in relatively compact regions of high density \citep{chung2021}. The 3$-$2 transitions of $\ceo$ and $\tco$ show similar emission distributions to their 1$-$0 transitions, and they are useful to trace dense filament regions and their diffuse envelope regions, respectively. Hence, we used the $\tco~(3-2)$ emission to investigate the velocity fields covering diffuse gas over a larger area than that $\ceo~(3-2)$ emission traces.

We checked the $\tco~(3-2)$ spectra and found that most line profiles over the region have a single Gaussian shape, but that profiles near C5 (the easternmost core of F13S) and the northern part of C4 (the northernmost core of F13) present double peaks. Hence, we performed a single Gaussian fitting for each line profile over most of the region, but a double Gaussian fitting for the spectra showing double peaks near C5 and the northern part of C4. We selected the higher velocity component between the two fitted components at the northern part of C4 and the lower velocity component near C5, which are similar to the velocity field of F13 and F13S, respectively. The central velocity map of $\tco~(3-2)$ is displayed in the left panel of Figure~\ref{fig:vfields}, revealing a velocity field covering a broader region compared to $\ceo~(3-2)$. In particular, in the northwest region, we can observe a low-velocity filament that was not detected in $\ceo$. For F13, we can confirm the well-defined north-south velocity gradient seen in $\ceo$ as well as velocity gradients perpendicular to the major axis of the filament. 

%=========== FIGURE: B-vectors
\begin{figure*} \begin{center} %\epsscale{1.17}
\includegraphics[width=0.7\textwidth]{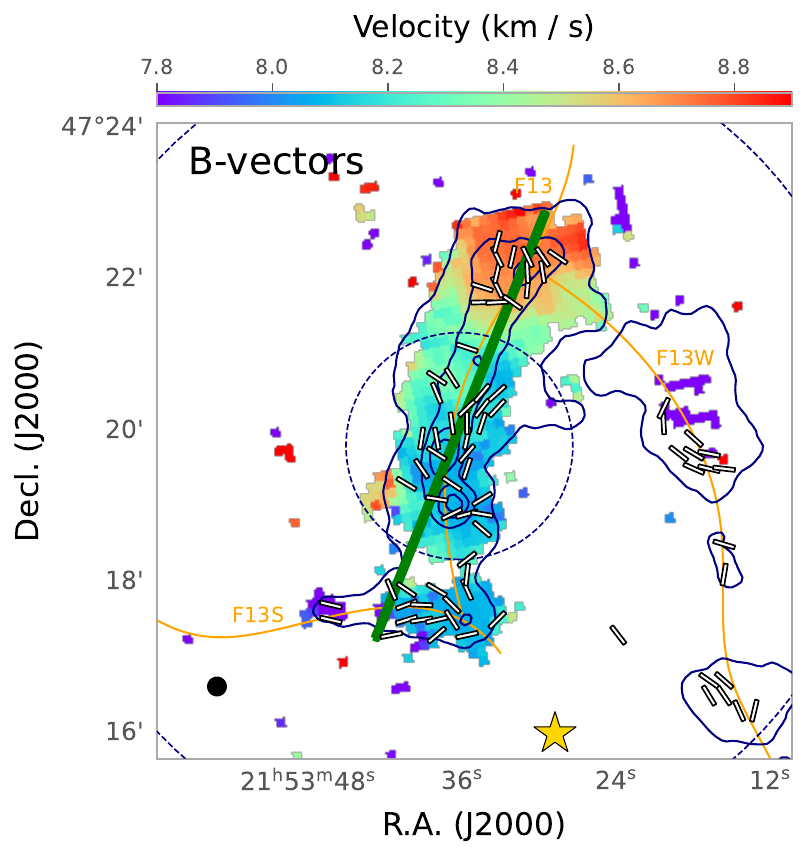} \end{center}
\caption{Magnetic field orientations on the $\ceo~(3-2)$ moment 1 map. The navy contours are the Herschel column density of 30, 70, 110, and 150$\times 10^{20}~ \rm cm^{-2}$. The selection criteria for magnetic field vectors are the same as in Figure~\ref{fig:polvectors}. All the B-field segments are shown with equal lengths to better display the B-field orientation. The green segment displays the large scale B-field orientation deduced from the Planck 353~GHz polarization vector. The yellow star at the south and the black circle at the bottom left corner represent the position of BD+46 and the POL-2 850~$\mu$m beam size of 14$\as$.1, respectively. \label{fig:bvectors}}
\end{figure*}
%=========== 

The transverse velocity fields at the positions of the cores are displayed in the right panels of Figure~\ref{fig:vfields}. The $\ceo$ emission is observed at small $r$ (distance from the skeleton) and shows velocity gradients of $\lesssim 1~\kms ~\rm pc^{-1}$. The $\tco$ data trace the velocity structure at larger $r$ compared to the $\ceo$ data, and it is quite intriguing that they show that the velocity decreases as $r$ decreases in all cores except for C4 core. The transverse velocity gradients range about 1 to 5~$\kms~\rm pc^{-1}$. This is quite larger than the velocity gradient along the major axis of the filament, but similar to the transverse velocity gradient shown in B211 filament of Taurus \citep[$\sim 2\kms~\rm pc^{-1}$;][]{palmeirim2013}. The V-shape (or $\Lambda$-shape) transverse velocity structures have been observed in several filaments, which is explained as the result of the ongoing compression due to the propagating shock fronts \citep[e.g.,][]{arzoumanian2018,bonne2020b,arzoumanian2022}.

%=========== FIGURE: B-vectors histo
\begin{figure} \begin{center} %\epsscale{1.17}
\includegraphics[width=0.49\textwidth]{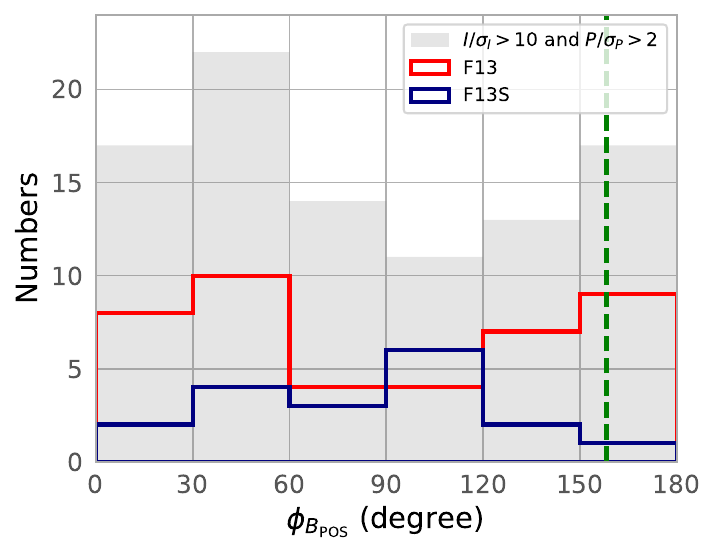} \end{center}
\caption{Histogram of the position angles for the magnetic field orientations presented in Figure~\ref{fig:bvectors}. The filled histogram is for all the B-field vectors with $I/\sigma_{\rm I} > 10$ and $P/\sigma_{\rm P}>2$ over the observed region, and the red and navy histograms are for those in F13 and F13S, respectively. The green dashed line indicates the Planck B-field orientation (see the green segment in Figure~\ref{fig:bvectors}). \label{fig:bv_histo}}
\end{figure}
%=========== 

We calculated the nonthermal velocity dispersion ($\sigma_{\rm NT}$) by extracting the thermal velocity dispersion ($\sigma_{\rm T}$) from the observed total velocity dispersion ($\sigma_{\rm obs}$):
\begin{equation}
\sigma_{\rm NT} = \sqrt{\sigma_{\rm obs}^{2} - \sigma_{\rm T}^{2}}, \label{eq:sigmaNT}
\end{equation}
where the observed total velocity dispersion was taken from the Gaussian fitting result. The thermal velocity dispersion of the observed molecule is
\begin{equation}
\sigma_{\rm T} = \sqrt{\frac{k_{\rm B} T}{\mu_{\rm obs} m_{\rm H}}},
\end{equation}
where $k_{\rm B}$ is the the Boltzmann constant, $T$ is the gas temperature, $\mu_{\rm obs}$ is the atomic weight of the observed molecule (30 for $\ceo$ and 29 for $\tco$), and $m_{\rm H}$ the hydrogen mass, respectively. We used the dust temperature obtained from Herschel continuum data for the gas temperature \citep{arzoumanian2011}. The resulting $\sigma_{\rm NT}$ is in a range of about $0.2 - 0.5~\kms$ with the mean value of $0.29\pm0.06~\kms$. 

The nonthermal velocity dispersion ($\sigma_{\rm NT}$) of $\tco$ and $\ceo$ in units of the sound speed ($c_{\rm s}$) are presented in Figure~\ref{fig:sigNT}. The nonthermal velocity dispersion of $\tco$ and $\ceo$ are transonic to supersonic ($c_{\rm s} \lesssim \sigma_{\rm NT} \sim 2c_{\rm s}$) and subsonic to transonic ($0.5c_{\rm s} \lesssim \sigma_{\rm NT} \lesssim 1.5c_{\rm s}$), respectively. There is no increase in nonthermal velocity dispersion in specific regions of the filament, and there is no correlation with the column density. Rather, it can be seen that $\sigma_{\rm NT}$ of $\tco$ is about 1.5 times $\sigma_{\rm NT}$ of $\ceo$. This larger $\sigma_{\rm NT}$ of $\tco$ can be caused by its larger optical depth than that of $\ceo$. In addition, $\tco$ has a lower critical density than $\ceo$, allowing it to trace regions with lower densities. Indeed, the width of $\tco$ in the spatial distribution is about twice that of $\ceo$, and assuming that their depths in line-of-sight are equal to their widths, $\sigma_{\rm NT}$ of $\tco$ will be about $\sqrt{2}$ times $\sigma_{\rm NT}$ of $\ceo$. Thus, the constant ratio of non-thermal velocity dispersions between $\tco$ and $\ceo$ appears to be attributed to these distribution differences. \\

\section{Analysis} \label{sec:anal}

\subsection{Magnetic Field} \label{ssec:bfield}

\subsubsection{Magnetic Field Geometry} \label{sssec:bfieldgeo} 

Figure~\ref{fig:bvectors} illustrates the magnetic field orientations in the region. The large green segment is the large scale magnetic field orientation obtained from the Planck 353~GHz polarization data\footnote{https://irsa.ipac.caltech.edu/applications/planck/}. We averaged the $Q$ and $U$ intensities over the F13 and F13S regions of $\nht \geq 30\times10^{20}~\rm cm^{-2}$ and measured the polarization angle, and deduced the magnetic field angle by rotating the polarization angle by 90 degrees. The Planck data show that the averaged magnetic field over the filament is parallel to the main direction of F13. We note here that the size of F13 is quite smaller than the FWHM size of Planck observation \citep[$\sim 5^{\prime}$ at 353~GHz; Planck Collaboration Int][]{planck35}.

The core-scale magnetic field vectors inferred by rotating the 850~$\mu$m polarization vectors by 90 degrees are presented with white segments in Figure~\ref{fig:bvectors}. The magnetic field vectors in F13S align mostly parallel to the directions of filament's skeleton. However, in F13, the magnetic field orientations appear to be more complex. At the northern and western parts of F13, the magnetic field vectors tend to align parallel to the direction of the filament's skeleton. Conversely, perpendicular B-field vectors are observed at the southern and eastern parts. In Figure~\ref{fig:bv_histo}, the position angle distribution of B-field vectors is presented. The B-field angle distribution of F13 peaks at about 45 and 165 degrees, the former attributed to the vectors near C1, while the latter is close to the mean field orientation observed with Planck. The histogram of B-field orientation of F13S peaks at $\sim$105 degrees, which is consistent with the main direction of F13S. 

The magnetic field orientation with respect to the filament axis have been reported to change in a way that B-fields are aligned parallel to the filament at low column density, but become perpendicular at high column density of $\nht \simeq 10-100\times 10^{20}~\rm cm^{-2}$ \citep[e.g., Planck Collaboration Int][]{planck35}. In the case of F13S, $\nht$ is $\lesssim 70 \times 10^{20}~\rm cm^{-2}$. $\nht$ range of F13 is much larger than that of F13S up to $\sim 170 \times 10^{20}~\rm cm^{-2}$. And, the magnetic field geometry is likely to be disordered at the high density regions of $\gtrsim 110 \times 10^{20}~\rm cm^{-2}$. 

F13S shows an increase in velocity along the east-west direction, which coincides with the filament's skeleton direction (see Figures~\ref{fig:longiv} and \ref{fig:vfields}). The magnetic field direction also aligns with this trend of velocity field. F13 also shows an increment in velocity from south to north, aligned to the main direction. At the southern region of F13, near C1, decrement in velocity is found from east to west (see the transverse velocity profile of C1 in Figure~\ref{fig:vfields}). The B-field orientations in the region tend to align east-west direction, which is close to the transverse velocity gradient. However, due to the low signal-to-noise ratio of the polarization vectors, we would like to tread cautiously in interpreting these results. \\

\subsubsection{Magnetic Field Strength} \label{sssec:bfieldstr}

We estimated the magnetic field strength using the modified Davis-Chandrasekhar-Fermi (DCF) method \citep{davis1951,chandrasekhar1953}. The DCF method assumes that the underlying magnetic field is uniform but distorted by the turbulence, and calculates the strength of the magnetic field in the molecular clouds using the angular dispersion of the magnetic field vectors ($\delta \phi$), velocity dispersion ($\sigma$), and number density of the gas ($\rho$) with the following equation \citep{crutcher2004etal}: 
\begin{align}
	B_{\rm pos} &= Q_{\rm c} \sqrt{4 \pi \rho} \frac{\sigma}{\delta \phi} \nonumber  \\
	&\approx 9.3 \sqrt{n_{\rm H_{2}}} \frac{\Delta v}{\delta \phi}, \label{eq:dcf}
\end{align}
where $Q_{\rm c}$ is the correction factor for the overestimation of the magnetic field strength due to the beam integration effect. $Q_{\rm c}$ of 0.5 was adopted from \citet{ostriker2001}. $n_{\rm H_{2}}$ is the number density of the molecular hydrogen in cm$^{-3}$, and $\Delta v = \sigma_{\rm NT} \sqrt{\rm 8 ln2}$ in $\kms$. 

%=========== FIGURE: bgsubtract
\begin{figure*} \includegraphics[width=1\textwidth]{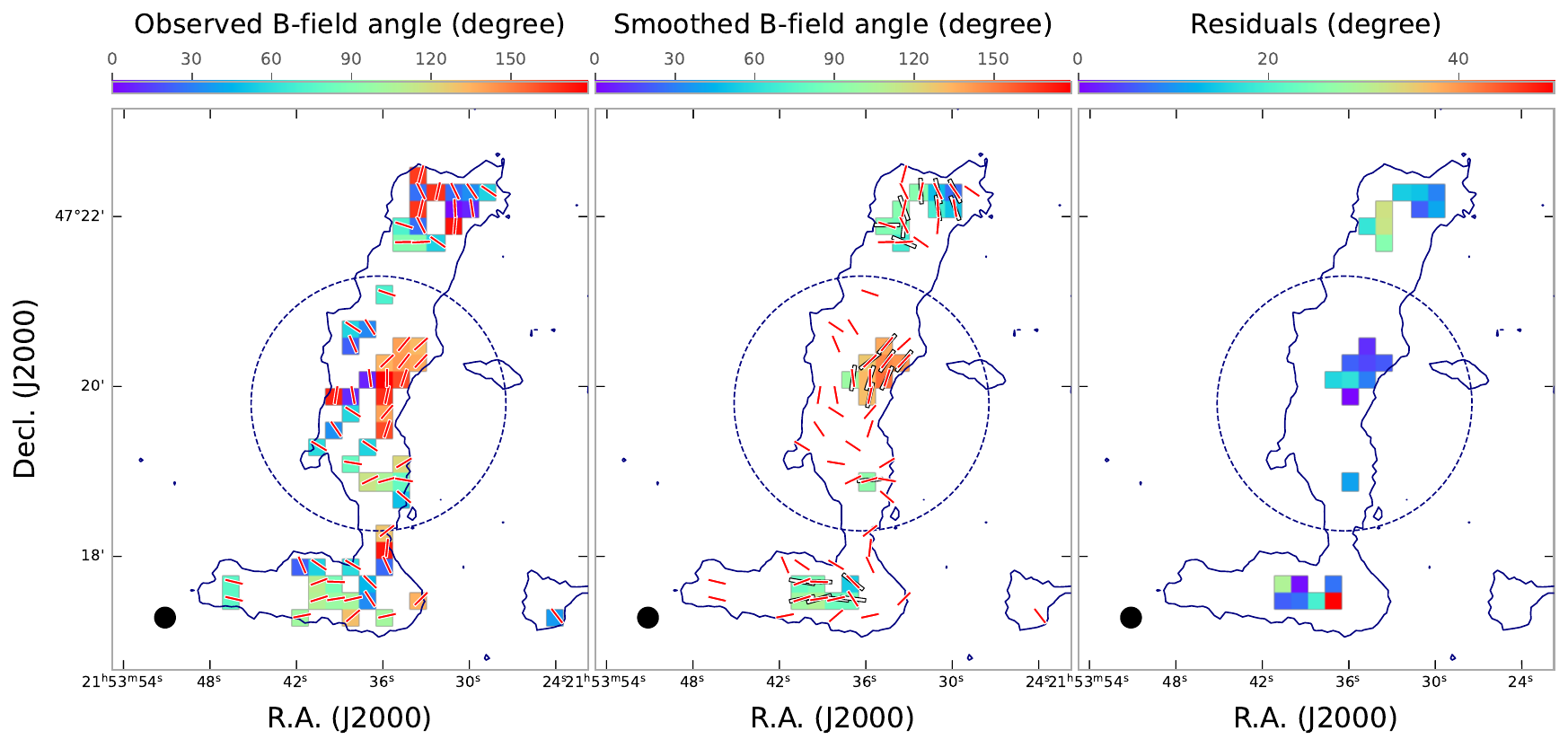} \caption{The magnetic field angle maps of the observed (left), smoothed (center), and residual (right). The orientations of the observed magnetic field vectors are drawn with red bars in the left and central panels and the smoothed B-field vectors are presented with white bars in the central panel. The data of smoothed and residual maps are shown only if the number of selected B-field vectors within a 3$\times$3 box is $\geq 5$. The dashed and solid circles indicate the 3$^{\prime}$ central region and 14$\as$.1 JCMT beam size at 850$\mu$m. \label{fig:um}}
\end{figure*}
%=========== 

The mean H$_2$ volume density along the crest ($\bar n_{\rm H_{2}}^{0}$) and $\Delta v$ measured from the nonthermal velocity dispersion given in Table~\ref{tbl:fila} were used. 

To estimate the angular dispersion of magnetic field vectors ($\delta \phi$), we adopted two independent methods. One is the unsharp-masking method \citep{pattle2017} and the other is the structure function \citep{hildebrand2009}. The unsharp-masking method measures the angular dispersion of the magnetic field distorted by the turbulence motions by removing the underlying magnetic field geometry \citep{pattle2017}. To obtain the large-scale background magnetic field structure, the observed magnetic field map is smoothed with an $n \times n$ pixel boxcar filter, where $n$ is an odd number. After obtaining the large-scale background magnetic field structure, the smoothed map is subtracted from the original map. Finally, the angular dispersion is measured from the residual map. We applied $3 \times 3$ pixel boxcar filter to the observed magnetic field map. 

%=========== FIGURE: structure function
\begin{figure}[b!] %\epsscale{1.17}
\plotone{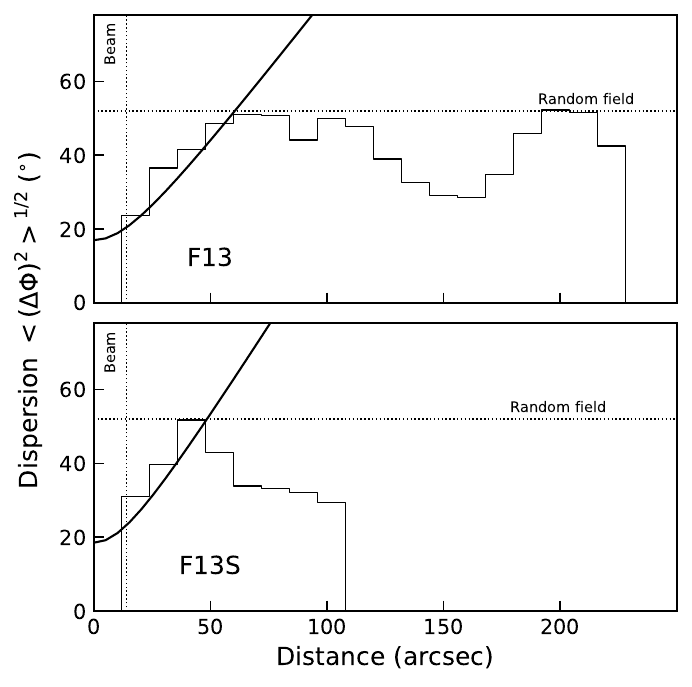}
\caption{The angular dispersion function ($\langle(\Delta \Phi)^{2}\rangle^{1/2}$) is shown for F13 (top) and F13S (bottom). The solid thick curve represents the best-fit model. The intercept of the fit at the zero point determines the contribution of turbulence to the total angular dispersion. The vertical dotted line indicates the beam size of POL-2 at the 850~$\mu$m wavelength, which is 14$^{\as}$.1. The horizontal dotted line represents the expected $<(\Delta \Phi)^{2}>^{1/2}$ for a random field, which is 52$^{\circ}$. \label{fig:sf}}
\end{figure}
%=========== 

Figure~\ref{fig:um} shows the position angle map of the observed magnetic field vectors (left), smoothed position angle map (middle), and the residual map (right). To avoid underestimating $\delta \phi$ due to the non-detected pixels and the small number of pixels in the boxcar filter, we used the residual values only when the number of data points in the 3$\times$3 boxcar filter is at least five in the calculation of $\delta \phi$. The obtained angular dispersions are $\delta \phi =$ 9.1$\pm$2.4 and 16.0$\pm$2.6 degrees for F13 and F13S, respectively.

\begin{deluxetable*}{lccccc} 
\input{tbl_bfield.tex}
\end{deluxetable*}

The second method we have used to determine the angular dispersion of the magnetic field vectors is the structure function model, which is `designed to avoid inaccurate estimates of turbulent dispersion due to a large-scale, nonturbulent field structure' \citep{hildebrand2009}. In the model, the structure function of the angle difference in a map is calculated as the following equation:
\begin{equation}
\langle\Delta \Phi^{2}(l)\rangle \equiv \frac{1}{N(l)} \sum\limits_{i=1}^{N(l)} [\Phi(x) - \Phi(x+l)]^{2} ,\label{eq:sf}
\end{equation}
where $\Phi(x)$ and $\Delta \Phi (l) \equiv \Phi(x) - \Phi(x+l)$ are the angle at the position $x$ and the angle difference between the vectors with separation $l$. $N(l)$ is the number of pairs of the vectors. The magnetic field is considered to consist of two components: a large-scale magnetic field and a turbulence scale magnetic field affected by the turbulence. As the scale $l$ increases within the range $0 \leq l \ll d$, where $d$ is the scale of the large-scale structured magnetic field, the contribution of the large-scale magnetic field to the dispersion function is expected to increase nearly linearly. The impact of turbulence on the magnetic fields can be described as follows: (1) At very small scales ($l \rightarrow 0$), the effect of turbulence is almost negligible or close to zero, (2) the turbulence has the most significant influence at scales approximately equal to the turbulent scale ($\delta$), and (3) for scales larger than the turbulent scale ($l > \delta$), the effect of turbulence remains constant. Then, the Equation~\ref{eq:sf} can be presented as follows \citep{hildebrand2009}:
\begin{equation}
	\langle\Delta \Phi^{2}(l)\rangle_{\rm tot} \simeq b^{2} + m^{2}l^{2} + \sigma_{\rm M}^{2}(l), \label{eq:sf2}
\end{equation}
where $\langle\Delta \Phi^{2}(l)\rangle_{\rm tot}$ is the square of the total measured dispersion function. The term $b$ represents the constant turbulent contribution to the angular dispersion within the range $\delta < l < d$. The parameter $m$ characterizes the linearly increasing contribution of the large scale magnetic field. $\sigma_{\rm M}^{2}(l)$ accounts for the correction term due to the measurement uncertainty when dealing with real data. 

Figure~\ref{fig:sf} illustrates the square root of corrected angular dispersion ($\langle\Delta \Phi^{2}(l)\rangle_{\rm tot} - \sigma_{\rm M}^{2}(l)$) as a function of distance $l$. The data have been divided into distance bins with separations corresponding to the pixel size of 12$\as$. We performed best-fit analyses using the first three data points to ensure that the condition $l \ll d$ is satisfied. The parameter $b^{2}$ was obtained from the least square fitting of the relation, resulting in estimated values of $b$ for F13 and F13S as $17.0$ and $18.5$, respectively. The corresponding angular dispersions $\delta \phi=\sqrt{b^{2}/2}$ to be applied to the modified DCF method are $12.0\pm6.2$ and $13.1\pm2.1$ degrees for F13 and F13S, respectively. 

Table~\ref{tbl:bfields} presents the applied $\bar{n}_{\rm H{2}}$, $\Delta v$, $\delta \phi$, and measured magnetic field strengths for F13 and F13S. The magnetic field strengths derived from the unsharp-masking method are 76$\pm$23 and 33$\pm$7 $\mu$G for F13 and F13S, respectively. Using the structure function method, the estimated magnetic field strengths are 58$\pm$31 and 40$\pm$9 $\mu$G for F13 and F13S, respectively. It is important to note that the values of $B_{\rm POS}$ obtained from both methods agree well with each other within the uncertainties involved. Hereafter, we will denote quantities obtained using the unsharp-masking method with $^{\rm UM}$ and those from the structure function method with $^{\rm SF}$. \\

\subsection{Magnetic Field, Gravity, and Turbulence} \label{ssec:bgt}

We estimated the mass-to-magnetic flux ratio ($\lambda$) and Alfv\'enic Mach number ($M_{\rm A}$) to investigate the significance of magnetic fields with respect to the gravity and turbulence. 

%=========== FIGURE: structure function
\begin{figure} \epsscale{1.17}
\plotone{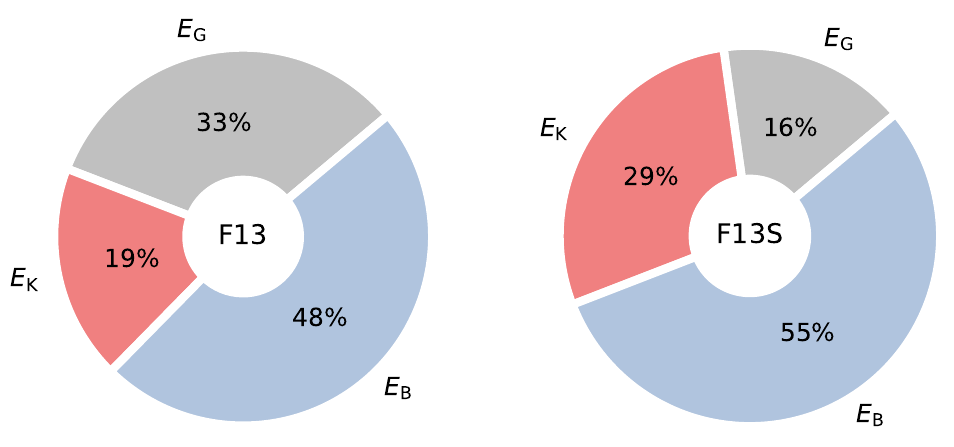}
\caption{Respective energy components, including gravitational energy ($E_{\rm G}$), kinematic energy ($E_{\rm K}$), and magnetic energy ($E_{\rm B}^{\rm SF}$) are depicted for F13 (left) and F13S (right) with their relative proportions represented by distinct colors: gray for $E_{\rm G}$, red for $E_{\rm K}$, and blue for $E_{\rm B}^{\rm SF}$. The values indicate the occupying percentage of each energy component. \label{fig:energies}}
\end{figure}
%=========== 

%=========== FIGURE: spatial B_pos
\begin{figure*} %\epsscale{1.17}
\includegraphics[width=0.99\textwidth]{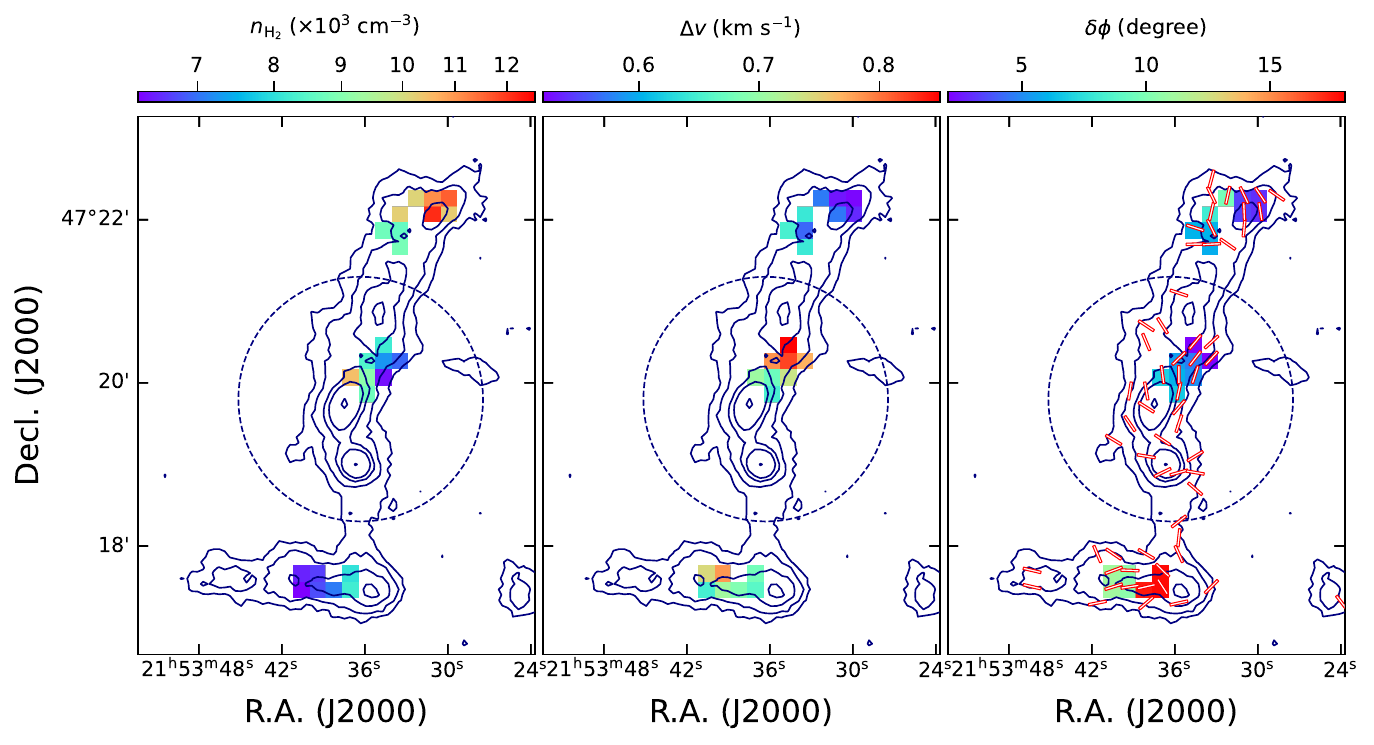} \caption{The spatial distributions of $n_{\rm H_{2}}$, $\Delta v$, and $\delta \phi$ with the 850~$\mu$m emission contours. The data is shown only if $\delta \phi$ is measured where the number of angle residual data points within a 3$\times$3 box is $\geq 3$. The observed magnetic field vectors with $I/\sigma_{I} \geq 10$ and $P/\sigma_{P} \geq 2$ are drawn in the $\delta \phi$ map.} \label{fig:spatial}
\end{figure*}
%=========== 

The mass-to-magnetic flux ratio ($\lambda_{\rm obs}$) is presented as:
\begin{equation}
	\lambda_{\rm obs} = \frac{(M / \Phi)_{\rm obs}}{(M / \Phi)_{\rm crit}}, \label{eq:m2mfr}
\end{equation}
where $(M/\Phi)_{\rm obs}$ and $(M/\Phi)_{\rm crit}$ are the observed and the critical mass-to-magnetic flux ratios, respectively. $(M/\Phi)_{\rm obs}$ is calculated with the following Equation:
\begin{equation}
	(M / \Phi)_{\rm obs} = \frac{\mu_{\rm H_{2}} m_{\rm H} N_{\rm H_{2}}}{B_{\rm pos}},
\end{equation}
where $\mu_{\rm H_{2}}$ and $N_{\rm H_{2}}$ are the mean molecular weight per hydrogen molecule of 2.8 and the H$_2$ column density. The critical mass-to-magnetic flux ratio is $(M/\Phi)_{\rm crit} = 1/(2 \pi \sqrt{G})$ \citep{nakano1978}.
Equation~\ref{eq:m2mfr} can be expressed following \citet{crutcher2004etal} as:
\begin{equation}
\lambda_{\rm obs} = 7.6 \times 10^{-21} \frac{N_{\rm H_{2}} [\rm cm^{-2}]}{B_{\rm pos} [\mu \rm G]}. \label{eq:lambda}
\end{equation}
The mean H$_{2}$ column density on the crest ($\bar N_{\rm H_{2}}^{0}$) was used for $N_{\rm H_{2}}$. The actual $\lambda$ value is assumed to be $\lambda_{\rm obs} / 3$ considering the statistical correction factor of 3 for the random inclination of the filament \citep{crutcher2004etal}. The criterion for the significance of magnetic fields is given in a way that if $\lambda$ is less than 1, then magnetic fields are expected to support the clouds, while $\lambda$ greater than 1 would mean the structure to be in a gravitational collapse.

For F13 and F13S, the values of $\lambda^{\rm UM}$ are $0.31 \pm 0.13$ and $0.27 \pm 0.09$, respectively. When using the structure function method, the corresponding values are $\lambda^{\rm SF}$ of $0.41 \pm 0.25$ and $0.22 \pm 0.08$, respectively. These results suggest that both F13 and F13S are likely supported by magnetic fields, as their $\lambda$ values are less than 1. 

We estimated the Alfv\'enic Mach number ($M_{\rm A}$) by:
\begin{equation}
	M_{\rm A} = \frac{\sigma_{\rm NT}}{V_{\rm A}}, \label{eq:mach}
\end{equation}
where $\sigma_{\rm NT}$ and $V_{\rm A}$ are the non-thermal velocity dispersion and the Alfv\'en velocity, respectively. $V_{\rm A}$ is defined as $V_{\rm A} = B / \sqrt{4 \pi \bar{\rho}}$ where $B$ and $\bar{\rho}$ are the total magnetic field strength and the mean density, respectively. The statistical average value of $B_{\rm pos}$, $(4/\pi)B_{\rm pos}$, was used for $B$ \citep{crutcher2004etal}, and the mean density was obtained from $\mu_{\rm H_{2}} m_{\rm H} \bar{n}_{\rm H_{2}}^{0}$. F13 and F13S have $V_{\rm A}^{\rm UM}$ of 1.21$\pm$0.36 and 0.79$\pm 0.16~\kms$ and $V_{\rm A}^{\rm SF}$ of 0.91$\pm$0.49 and 0.97$\pm 0.20~\kms$, respectively. The Alfv\'enic Mach numbers of the two filaments are less than 0.5, and hence F13 and F13S are both sub-Alfv\'enic indicating the magnetic fields dominate turbulence in the regions. 

%=========== FIGURE: spatial B_pos2
\begin{figure*} %\epsscale{1.17}
\includegraphics[width=0.99\textwidth]{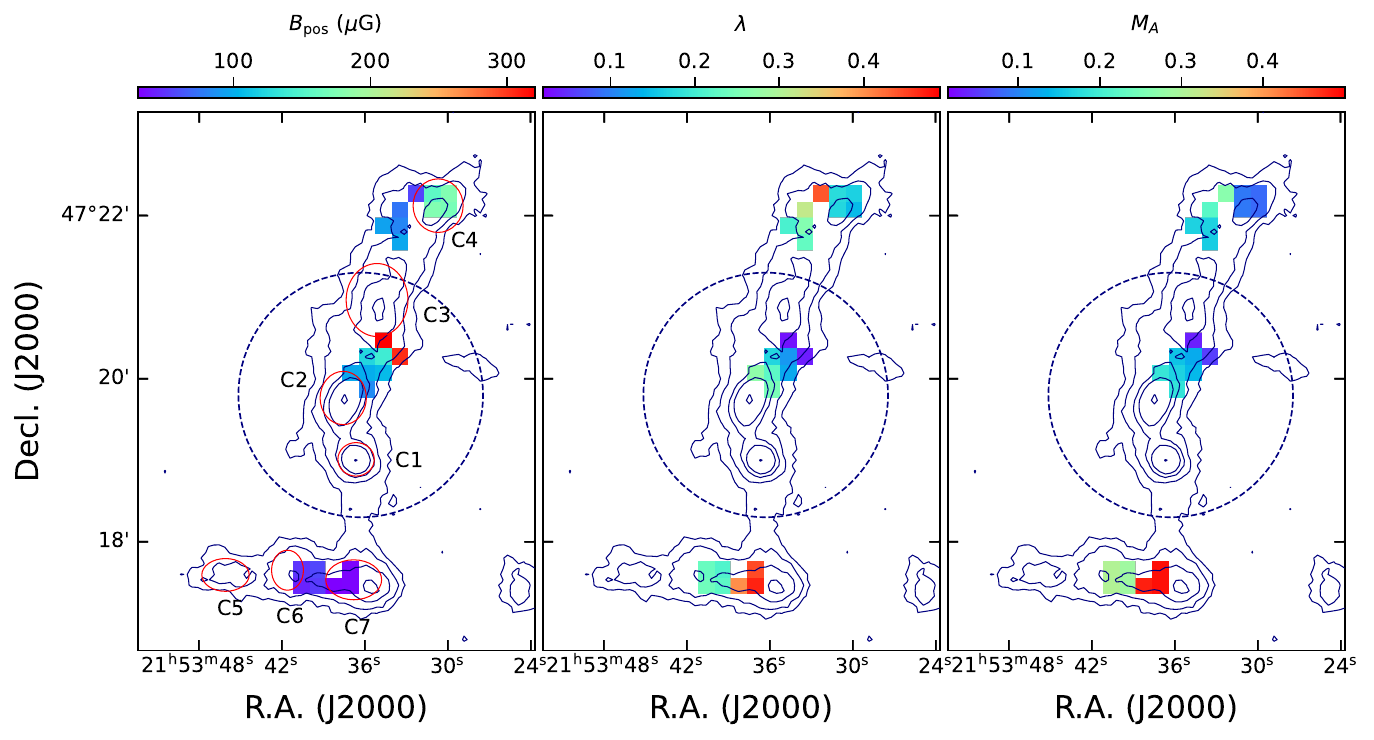} \caption{The spatial distributions of $B_{\rm POS}$, $\lambda$, and $M_{\rm A}$ with the 850~$\mu$m emission contours. The red ellipses in the $B_{\rm POS}$ map present the dense cores.} \label{fig:spatial2}
\end{figure*}
%=========== 

We calculated the total gravitational ($E_{\rm G}$), kinematic ($E_{\rm K}$), and magnetic field energies ($E_{\rm B}$) assuming the cylindrical geometry of F13 and F13S using the following equations \citep{fiege2000i}:
\begin{equation}
	E_{\rm G} = - \frac{GM^{2}}{L} \label{eq:eqegf},
\end{equation}
\begin{equation}
	E_{\rm K} = M \sigma_{\rm tot}^{2} ,
\end{equation}
and
\begin{equation}
	E_{\rm B} = \frac{1}{2} M V_{\rm A}^{2}, \label{eq:eb}
\end{equation}
where $M$ and $L$ are the mass and length of filament, respectively. $\sigma_{\rm tot}$ is the total velocity dispersion estimated by the equation:
\begin{equation}
	\sigma_{\rm tot} = \sqrt{\sigma_{\rm NT}^{2} + \frac{k_{\rm B} T}{\mu_{\rm p} m_{\rm H}}}, \label{eq:sigmatot}
\end{equation}
where $\mu_{\rm p}$=2.37 is the molecular weight of the mean free particle \citep{kauffmann2008}. 

The estimated quantities are tabulated in Table~\ref{tbl:bfields}. F13 exhibits significantly larger $E_{\rm G}$, $E_{\rm K}$, and $E_{\rm B}$, approximately 5 to 20 times greater than F13S. This difference can be attributed to F13's larger mass compared to F13S. In Figure~\ref{fig:energies}, the relative energy distributions are represented using donut diagrams, where $E_{\rm B}^{\rm SF}$ is employed to denote the energy portions. As depicted in the Figure, the largest energy contribution for both F13 and F13S comes from the magnetic field ($\gtrsim$50\%). The energy distributions of the two filaments which have the largest portion in the magnetic field energy agree to those of filaments which have quasi-periodically aligned dense cores \citep{chung2023}. Meanwhile, the portions of magnetic field energy in F13 and F13S are quite larger than those in the E- and W-hubs of the dark Streamer of IC~5146 \citep[$E_{\rm B}\sim 20$\%;][]{chung2022}. While the comparison is limited to a few filaments, this supports the proposal by \citet{chung2021} that, in the filaments of the Cocoon Nebula, thermal pressure and magnetic fields may be more crucial than turbulence compared to the filaments in the dark Streamer. \\

\subsection{B-field Distributions} \label{ssec:bdistri}

We obtained the distribution of magnetic field strengths by applying the modified DCF method in a pixel-by-pixel scale \citep[e.g.,][]{hwang2021}. A $\delta \phi$ map was obtained using the residual map of magnetic field angles in Figure~\ref{fig:um}. If the number of angle residuals within a 3$\times$3 box is $\geq 3$, the standard deviation of residual polarization angles was measured for $\delta \phi$. An H$_{2}$ number density map was estimated with Equation~\ref{eq:ndh2} and a $\Delta v$ map was made from the calculation of the nonthermal velocity dispersion as $\Delta v = \sigma_{\rm NT} \sqrt{8 \rm ln 2}$. The distribution maps of $n_{\rm H_{2}}$, $\Delta v$, and $\Delta \phi$ are presented in Figure~\ref{fig:spatial}. Due to the non-detected pixels mostly in dense regions, we have a total of 24 detected pixels in the map.

The spatial distribution of $B_{\rm pos}$ was obtained by using the $n_{\rm H_{2}}$, $\Delta v$, and $\Delta \phi$ distributions and is shown in Figure~\ref{fig:spatial2} with the mass-to-magnetic flux ratio map and the Alfv\'enic Mach number map obtained with the Equations~\ref{eq:lambda} and \ref{eq:mach}. The magnetic field strength mostly ranges from $\sim$30 to $200~\mu$G. %Although there are almost no data points in the high-density region, higher density regions tend to have higher magnetic field strength, especially at the directions between two dense cores, C2 and C3 ($\sim$140 and 110~$\mu$G, respectively). In the high density region like C4, such high magnetic field strength is also observed. 
The $\lambda$ and $M_{A}$ distributions show that the ranges are both between 0.2 and 0.5, indicating that the regions are magnetically subcritical and sub-Alfv\'enic throughout. However, although there are almost no data points in the high-density region, higher density regions tend to exhibit a higher mass-to-magnetic flux ratio, particularly in the directions toward two dense cores, C2 and C7. Hence, we do not rule out the possibility that the dense regions are magnetically supercritical, given that the uncertainties in $\lambda$ are large. \\

\section{Discussion} \label{sec:disc}

\begin{figure} \begin{center} %\epsscale{1.17}
\includegraphics[width=0.5\textwidth]{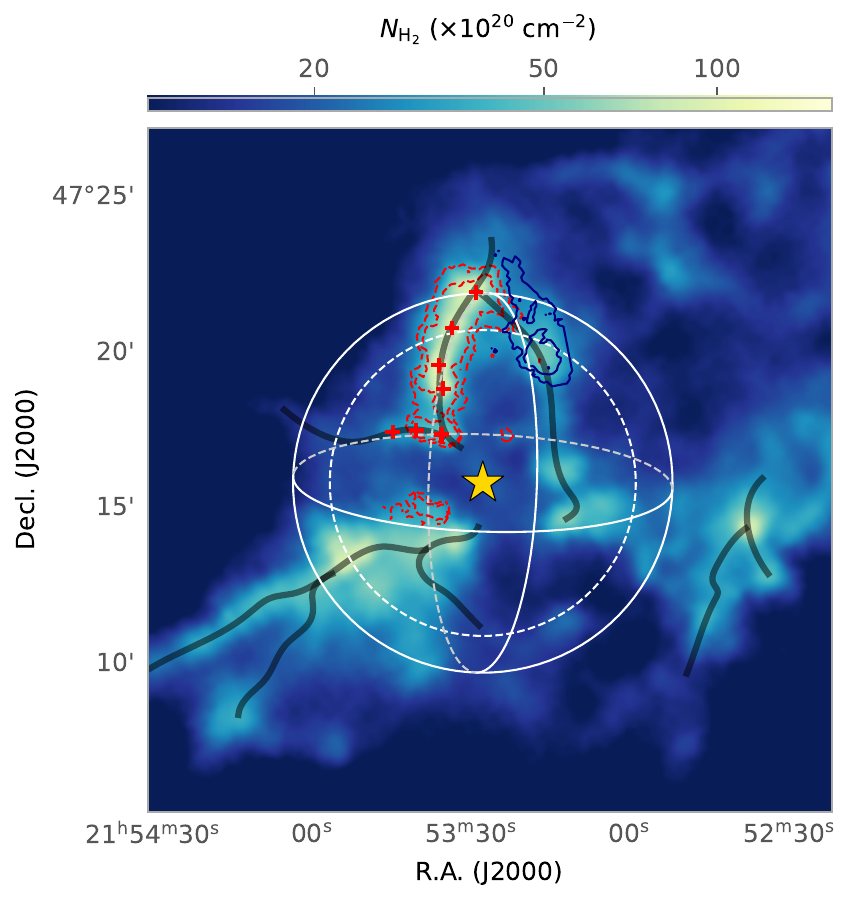} \end{center}
\caption{Imaginary picture of the Cocoon Nebula system. The H$_{2}$ column density map from Herschel is presented, and the filaments' skeletons are overlaid with thick black curves \citep{arzoumanian2011}. The positions of dense cores and BD+46 are displayed with red crosses and big yellow star, respectively. The blue and red $\tco$ components integrated over the velocity ranges of $7-8.4 ~\kms$ and $8-8.9~\kms$, respectively, are illustrated with solid navy and dotted red lines. The imaginary 3-dimensional sphere with a radius of 0.1$^{\circ}$ (identical to 1.44~pc at the assumed distance of 813~pc) is shown in white color. The white dashed circle indicates the size of {\sc Hii} region \citep[radius of 1.2~pc at the distance of 813~pc;][]{roger1982}. \label{fig:cocoon3d}}
\end{figure}

The basic picture of the Cocoon Nebula given by the {\sc Hi} observations is a blister {\sc Hii} region formed by a B0 V single star BD+46 in front of the parent molecular cloud \citep{roger1982}. It is reported that the maximum radius of {\sc Hii} region is 1.2~pc \citep{roger1982} and the {\sc Hi} to H$_{2}$ transition is occurred at the distance of 1.44~pc from BD+46 \citep{annestad2003} assuming the distance of 813~pc to the Cocoon Nebula. F13 and F13S are located behind the {\sc Hii} region formed by BD+46 \citep{roger1982}. Figure~\ref{fig:cocoon3d} presents the Herschel column density map with an imaginary 3-dimensional sphere. The skeletons of filaments in the region are overlaid, and the redshifted ($v_{\rm lsr} \sim 8.0 - 8.9~ \kms$) and blueshifted ($v_{\rm lsr} \sim 7.0 - 8.4~ \kms$) $\tco$ emissions are represented with red and blue contours. The {\sc Hii} region and the {\sc Hi} to H$_{2}$ transition region are depicted with the white dashed circle and solid sphere in Figure~\ref{fig:cocoon3d}, respectively. As shown in the Figure, F13 and F13S likely reside on the imaginary sphere with a radius of 1.44~pc. In this section, we will discuss the dust polarization properties, filament formation, and core formation using this picture of Cocoon Nebula. \\

%=========== FIGURE: 3d cocoon
%=========== 

\subsection{The radiation field of BD+46 and the dust polarization fraction} \label{ssec:pi}

%=========== FIGURE: distance - pi relation
\begin{figure*} %\epsscale{1.17}
\includegraphics[width=0.95\textwidth]{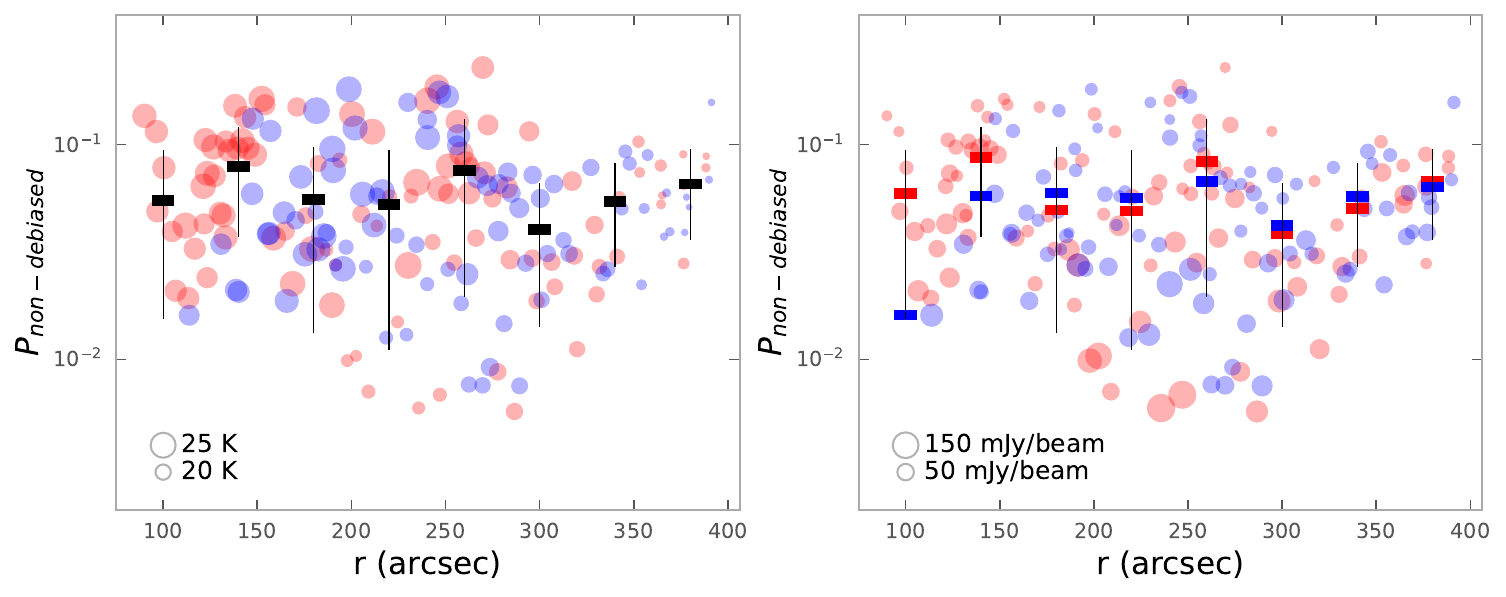}
\caption{The polarization fraction as a function of distance from BD+46. The two panels show the same distribution with the marker sizes scaled by the dust temperature in the left panel and the Stokes $I$ intensity in the right panel. The red and blue color indicate the position of vectors in the filaments, i.e., the red and blue are for the vectors close to and far from BD+46. The black thick horizontal bar in the left panel is the average of non-debiased $P$ in every 40$\as$ bin, and the red and blue horizontal bars in the right panel are those for the vectors close to and far from BD+46, respectively. \label{fig:dpf}}
\end{figure*}
%=========== 

\citet{ngoc2021} found that the California LkH$\alpha$~101 region surrounding an early B star, LkH$\alpha$~101, has the single power-law slope of $\alpha = 0.82\pm0.03$. They interpreted the rapid decrease of $P$ as $I$ increases is due to the mixing effect of the magnetic field tangling and grain alignment with rotational disruption by radiative torques. In the south of F13S, there is a B0 V single star, BD+46. $\alpha$ of the single power-law fit over the F13 region is $0.82\pm0.09$ estimated with the vectors of $I/\sigma_{I} \geq 10$ and $P/\sigma_{P} \geq 3$, and it is quite similar value to that of LkH$\alpha$~101.  

To examine whether the $\alpha$ of F13 region is related with the radiation field of BD+46, we divided the filament into two near and far regions from BD+46 with respect to the skeleton of the filaments. We applied the Ricean-mean model to those vectors of near and far regions separately. In Figure~\ref{fig:ipf}, the vectors of near and far regions are colored with red and blue, respectively, and their best fit results are drawn with the same color code. In Table~\ref{tbl:alpha}, the best-fit results are tabulated. Shown is that the vectors of near region to BD+46 have a slightly steeper slope ($\alpha = 0.73\pm0.07$) than the vectors in the far region have ($\alpha = 0.71\pm0.13$). However, the difference between them is smaller than their uncertainties. 

Figure~\ref{fig:dpf} presents the polarization fraction ($P$) plotted against the projected distance from BD+46. No significant correlation between distance and $P$ is observed. However, for distances smaller than 160$\as$, the closer vectors show a larger mean $P$ compared to the far vectors. This could serve as indirect evidence that the radiation field of BD+46 influences the dust alignment efficiency in the region, and consequently, the polarization fraction. It is crucial to note that there is only one far vector in the bin between $80\as < r \leq 120\as$, and thus, we cannot rule out the possibility of bias due to the small sample size. Above all, if F13 and F13S are indeed located on the surface of a sphere with a radius of 1.44 pc, as depicted in Figure~\ref{fig:cocoon3d}, it would provide a compelling explanation for the lack of correlation between the projected distance and the degree of polarization. \\

\subsection{Possible Formation of F13 by Radiation Shock Front of BD+46} \label{ssec:fform}

%=========== FIGURE: schematic figure of formation
\begin{figure*} \begin{center} %\epsscale{1.17}
\includegraphics[width=0.99\textwidth]{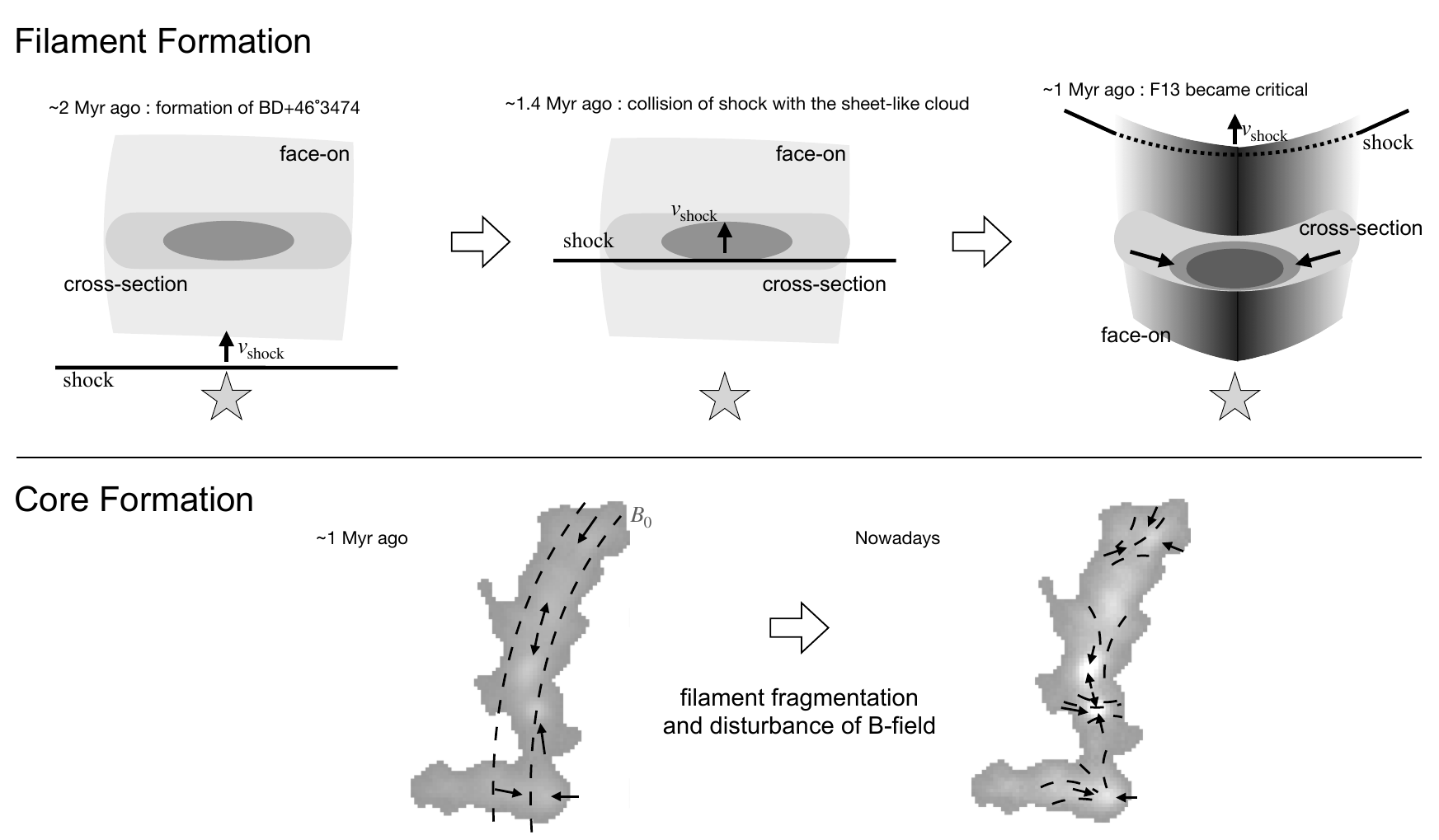} \end{center}
\caption{Schematic figure of the filament's formation and evolution. {\it Top:} Molecular cloud before and after the collision with the radiation shock front. The face-on view and cross section of the molecular cloud are colored with the gray tone, while the dense filament formed through the collision is shaded in a deep tone. The direction of shock front and the mass flow from the less dense ambient cloud to dense filament are represented with arrows. {\it Bottom:} Core formation by the filament fragmentation. The magnetic field and mass flow along the filament are presented with black dashed curves and arrows. \label{fig:fform}}
\end{figure*}
%=========== 

Theoretically, filaments can form through self-gravitational fragmentation within a sheet-like cloud created by shock compression \citep[e.g.,][]{tomisaka1983,nagai1998}. This compression is induced by feedback from massive stars, cloud-cloud collisions, galactic spiral shocks, as well as from expanding \textsc{Hii} regions. In a filament formed through shock compression within a sheet-like cloud, a notable characteristic is the V-shape (or $\Lambda$-shape) transverse velocity structure, which is attributed to mass flows induced along magnetic field lines or by self-gravity \citep[e.g.,][]{arzoumanian2018,inoue2018,abe2021}. In case that the ordered magnetic field causes mass flow from the natal cloud and formation of filament, the major direction of filament becomes perpendicular to the magnetic field direction \citep{inoue2013,inoue2018}. However, the large scale magnetic field direction deduced from the Planck polarization data is almost parallel to the major direction of F13, and the high resolution B-field vectors from the POL-2 polarization data are mostly longitudinal, too (see Figure~\ref{fig:bvectors}). Thus, the velocity structure of F13 and F13S is not likely induced by the magnetic field lines.

When the initial magnetic field strength is smaller than 200~$\mu$G, a sheet compressed via the {\sc Hii} region of a young OB star becomes unstable, leading to the formation of a dense filament \citep{tomisaka1983}. If the thickness of the layer is smaller than the pressure scale height, the major orientation of the filament aligns parallel to the magnetic field lines \citep{nagai1998}. The measured magnetic field strength of F13 ($\lesssim 140 \mu$G; see Table~\ref{tbl:bfields} and Figure~\ref{fig:spatial2}), along with the observed parallel magnetic field direction (Figure~\ref{fig:bvectors}), suggests the possibility that it may have originated from a thin shock-compressed layer around the {\sc Hii} region created by BD+46 with a weak magnetic field. Moreover, \citet{abe2021} conducted isothermal magnetohydrodynamics simulations to explore filament formation, specifically examining the effects of shock velocity and turbulence. In the fast shock mode (with a shock velocity of $\sim 7~\kms$), they found that filaments form regardless of the presence of turbulence and self-gravity. Conversely, in the slow shock mode with a shock velocity of $\sim 2.5~ \kms$ and low turbulence, self-gravity likely plays a more significant role. It is worth noting that \citet{roger1982} reported an expanding velocity of $\sim 2.3 ~\kms$ for the \textsc{Hi} gas in the Cocoon Nebula, with the ionization front on the far side moving at $2 ~\kms$. Consequently, it is presumed that F13 and F13S were formed in a low-shock environment, in which case self-gravity would have played a significant role. 

Figure~\ref{fig:fform} shows the formation process of F13 schematically. The age of BD+46 is estimated to be approximately 2 million years \citep{garcia2014}, so it can be inferred that the shockwave originated from BD+46 around 2 Myr ago. Assuming a shock velocity of $2.3~\kms$ and a distance of 1.44~pc between BD+46 and F13, it would take $6\times 10^{5}$~years for the shock to reach the parent molecular cloud of F13. The nonthermal velocity dispersion ($\sigma_{\rm NT}$) is used as indirect evidence of the turbulent motion of gas induced by shocks, accompanied by velocity variations \citep[e.g.,][]{pineda2010,kim2022}. The gas in the region to where the shock front reach would become more turbulent and have larger $\sigma_{\rm NT}$ values. However, there is no apparent correlation between the nonthermal velocity dispersion and the radiation shock effects (Figure~\ref{fig:sigNT}). The absence of any discernible impact from radiation shocks in the nonthermal velocity dispersion distribution is likely due to that, as mentioned above, the radiation shock front passed through approximately 1.4~Myr ago.

The shockwave might compress the sheet, causing the compressed region to become denser. Subsequently, condensation due to self-gravity would have occurred. The V-shaped velocity structure seen in Figure~\ref{fig:vfields} could be attributed to the accretion of material that occurred during the formation process of F13. The transverse velocity differences at the filament crest and the radial distance of 0.2~pc of F13 (0.1~pc of F13S) as shown in Figure~\ref{fig:vfields} are $\sim 0.2 - 0.7 ~\kms$ ($\sim 0.1 - 0.5~\kms$ for F13S). Considering these transverse velocity differences, the infall velocity from the natal cloud, $v_{\rm inf}$, is expected to be $0.4\pm0.1~\kms$ and $0.3\pm0.1~\kms$ for F13 and F13S, respectively. The mass accretion rate per unit length onto the filament ($\dot{M}_{\rm line}$) from the natal cloud can be measured with the following Equation \citep[e.g.,][]{palmeirim2013,arzoumanian2022}: 
%\begin{equation} 	\dot{M} = \int_{r_{1}}^{r_{2}} 2 \times L \times W \times~\rho(r) \times \nabla v_{\rm t}(r) ~dr, \end{equation} 
\begin{equation}
 	\dot{M}_{\rm line} = \rho(R) \times v_{\rm inf} \times 2 \pi R, 
\end{equation} 
\noindent where $\rho(R)$ is the density at the infall radius $R$ (0.2~pc for F13 and 0.1~pc for F13S). $\rho(R)$ is estimated to be $n_{\rm H_{2}}=2\times 10^{3}~\rm cm^{-3}$ from the column density at $R$ ($2 \times 10^{21}~\rm cm^{-2}$) divided by F13's width (0.32~pc), and the mean molecular weight of 2.8 is used. The measured mass accretion rate per unit length is $\sim 70 \pm 30 ~M_{\odot}~\rm pc^{-1}~ Myr^{-1}$ for F13 and $\sim 30 \pm 10 ~M_{\odot}~\rm pc^{-1}~ Myr^{-1}$ for F13S. Then, it would take $9 \pm 5\times 10^{5}$~yr and $8 \pm 4 \times 10^{5}$~yr for F13 and F13S to have current mass per unit lengths of $67~M_{\odot}~\rm pc^{-1}$ and $22~M_{\odot}~\rm pc^{-1}$ at the current accretion rates. The thermal critical line masses of F13 and F13S are $27~M_{\odot}~\rm pc^{-1}$ and $33 ~M_{\odot}~\rm pc^{-1}$, respectively, and they may become critical in $4 \pm 1 \times 10^{5}$~yr for F13 and $12 \pm 5 \times 10^{5}$~yr for F13S. This suggests that since the formation of BD+46 occurred approximately 2 Myr ago, the radiation shock front collided with the sheet-like cloud around 1.4 Myr ago. F13 increased its line mass through infall mass driven by self-gravity, reaching a critical state around 1 Myr ago, potentially leading to the formation of dense cores. In the case of F13S, it became critical approximately 0.2 Myr ago. This indicates that a considerable period has elapsed since the formation of F13, enabling the formation of dense cores within the filament up to the present moment. In contrast, this has not been the case for F13S, possibly leading to less distinct fragmentation features due to a limited time for evolution. \\

\subsection{Filament Fragmentation and Core Formation} \label{ssec:ffcf}

After the formation of F13 and F13S through shock compression and mass infall due to self-gravity, it is likely that dense cores were generated through the fragmentation process. F13 and F13S have four and three dense cores, respectively, and the cores are well aligned on their parental filament's crest. The mean projected separations of cores in F13 and F13S are 0.27$\pm$0.08 and 0.20$\pm$0.05~pc, respectively. %These values are similar to those of filaments' widths, and quite smaller than the core separation ($\sim 0.9-1.3$ pc) expected by the classical linear fragmentation model \citep[e.g.,][]{inutsuka1992}. The theoretical model assumes an isothermal, infinitely long cylindrical structure, and its filament fragmentation occurs through gravitational perturbations with critical wavelengths equal to 2 times the diameter of the filament. Then the filaments may possess cores with a regular spacing of 4 times the diameter of the filament, representing the fastest growing mode of fragmentation. However, other factors such as turbulence, accreting flows, and/or magnetic fields may influence the fragmentation process, resulting in variations in core separations according to their respective environments \citep[e.g.,][]{fiege2000i,clarke2016,hanawa2017}. Observationally, filaments having core separations different from the model's predicted values are also observed, further supporting the idea that these factors play a significant role in shaping cores in filament structures \citep[e.g.,][]{tafalla2015,zhang2020,chung2023}. 

It is in broad agreement that the periodically aligned cores are the fragments of cylindrical filaments via the gravitational instability, which was firstly proposed by \citet{chandrasekhar1953b} and refined later in a few studies \citep[see \S4 of ][]{jackson2010}. The theoretical model assumes an isothermal, infinitely long cylindrical structure, and its fragmentation occurs through gravitational perturbations with critical wavelengths. Then the filaments may possess cores with a regular spacing, representing the fastest growing mode of fragmentation. In the pressure confined-models where the isothermal cylinder with the external pressure by an ambient medium is assumed, the wavelength of the fastest growing mode, $\lambda_{\rm max}$, is proposed to be $\sim 5\times$filament's width \citep{nagasawa1987,fischera2012}. The mean core separations in F13 and F13S are similar to those of filaments' widths, and quite smaller than that expected by the model.

The disagreement between the theoretical prediction and observational core spacings could be caused by their different density profiles from that assumed in the theoretical model. \citet{nagasawa1987} stated the case in which the isothermal scale height ($H=c_{\rm s}/\sqrt{4 \pi G \rho_{c}}$) of filament is much smaller than the filament's radius ($H \ll R$). In this case, $\lambda_{\rm max}$ equals $22\times H$. $H$ estimated with the mean dust temperature and central density in Table~\ref{tbl:fila} are 0.012~pc and 0.016~pc for F13 and F13S, respectively, and their radii are larger than $\sim$14 and 7 times the isothermal scale heights, respectively. Thus, $\lambda_{\rm max}=22\times H = 0.26$~pc for F13 closely matches the observed core spacing, while the mean core separation of F13S remains smaller than $\lambda_{\rm max}=0.35$~pc. Therefore, we can infer that the fragmentation of thermally supported F13 aligns with the cylindrical collapse model.

Meanwhile, the cores aligned with shorter core spacings than the theoretical prediction are commonly observed \citep[e.g.,][]{tafalla2015,zhang2020,shimajiri2023}. This is likely because the idealized equilibrium model cannot fully describe the real filaments, and the short spacing of the dense cores along the filament can be caused by a combination of the bent geometry of the filament, mass accretion from the natal cloud into the filament, and the magnetic field geometry to the filament \citep[e.g.,][]{fiege2000i,clarke2016,hanawa2017}. We find some supporting points for this in our observations. For example, F13 shows possible bending structures at least around C3 and C4 from the longitudinal velocity variation shown in Figure~\ref{fig:longiv}. In addition, the $\tco$ velocity field suggests that there may be a mass accretion into F13 from the natal surrounding cloud. Moreover, the magnetic field vectors obtained from our POL-2 high resolution observations are mostly parallel to the F13S's direction and those in F13 are also parallel to the filament's main direction except near the high density core region (i.e., C1 and C4). %Overall our observations suggest that above mentioned three factors may work for the fragmentation of filament into cores in smaller separation than the expected 4$\times$filament width. 

In Figure~\ref{fig:fform}, we schematically illustrated the formation process of dense cores within the filaments. Approximately 1~Myr ago, F13 and F13S formed, accompanied by a magnetic field oriented in the north-south direction. As the filament becomes gravitationally unstable, the fragmentation occurred, and during this process, influenced by the parallel magnetic field, the fragmentation likely happened at intervals shorter than four times the filament width. Additionally, as dense cores formed, the orientation of the magnetic field would have changed due to the influence of gravity and mass flow, resulting in the current distorted magnetic field configuration near the core regions.  \\

\section{Summary} \label{sec:summ}

We have carried out observations of 850~$\mu$m dust polarization and $\ceo$ and $\tco~(3-2)$ molecular lines toward the filaments (F13 and F13S) in the Cocoon Nebula (IC~5146) with the JCMT POL-2 and HARP instruments. The main results of our analysis are summarized below.

\begin{enumerate}
\item By using the column density map and the filament skeletons provided by the Herschel Gould Belt Survey \citep[HGBS;][]{andre2010,arzoumanian2011}, we measured the length, width, H$_{2}$ number density, and mass per unit length of F13 and F13S, finding that F13 and F13S are both thermally supercritical. We identified dense cores using the {\sc FellWalker} algorithm, and found four and three dense cores along the crests of F13 and F13S, respectively. The mean projected separation between the cores is close to the filament's width. 
\item We found that the polarization fraction has a dependency on 850~$\mu$m intensity in a single power-law ($P \sim I^{-\alpha}$) whose fit index $\alpha$ for the data of the selection criteria of $I/\sigma_{I} \geq10$ and $P/\sigma_{P} \geq 3$ is given as $0.82 \pm 0.09$. The polarization fraction data were also found to be fitted well with a Ricean-mean model. Using a Ricean-mean model with vectors having $I/\sigma_{I} \geq 10$, we obtain $\alpha = 0.72 \pm 0.07$. This indicates that the dust grains are well aligned inside the cloud, while the degree of such alignment possibly decreases at the dense region. We examined the effect of a B0 V star BD+46 on the dust alignment efficiency in the far side and near side from the star, finding that within the distance of $< 160\as$ from BD+46, there is a tendency of the near side vectors to have higher mean polarization fraction than the far side vectors. However, this might be attributed to the limited sample size, and F13 and F13S are arranged on a sphere, ensuring that the radiation field from BD+46 uniformly affects the entire region of the filaments.
\item The magnetic field strengths and their related physical quantities of two filaments F13 and F13S were derived to discuss their physical status. The magnetic field strengths were estimated using the modified DCF method, which are $B_{\rm pos}^{\rm SF} = 58 \pm 31$ for F13 and $40 \pm 9~\mu$G for F13S. The mass-to-magnetic flux ratio ($\lambda$) and the Alfv\'enic Mach number ($M_{\rm A}$) were calculated, finding that both filaments are magnetically subcritical and sub-Alfv\'enic. However, the magnetic field strength map and mass-to-magnetic flux ratio map show higher mass-to-magnetic flux ratio in denser regions, suggesting that these areas may be magnetically supercritical. We also estimated the gravitational ($E_{\rm G}$), kinematic ($E_{\rm K}$), and magnetic energies ($E_{\rm B}$) of F13 and F13S, to compare them together, finding that both filaments have the largest fraction in $E_{\rm B}$ ($\simeq 50\%$). The portions of $E_{\rm B}$ in F13 and F13S are quite close to that in the filaments of L1478. However, they are much higher than the portions of $E_{\rm B}$ in the eastern- and western hubs of the dark Streamer of IC~5146, indicating that the importance of magnetic fields is more significant in the Cocoon region than in the dark Streamer. 
\item We compared the $\ceo~(3-2)$ velocity with the $\nht$ profile along the filament's skeleton. We found that the velocity and density oscillations have approximately the same wavelength $\lambda$, but with a $\lambda/4$ shift between them around C1, C2, and C7. This suggests that the cores have formed by the fragmentation of a gravitationally unstable filament. The core spacings are found to be shorter than the value of $\sim 5 \times$filament width predicted in the cylindrical collapse model, but that of F13 is well matched with the prediction ($\sim 22 \times H$) in the case of isothermal scale height ($H$) is much smaller than the filament's radius implying that the cylindrical collapse model works well in the F13 fragmentation. Besides, the core spacings of F13 and F13S may also be influenced by gas accretion from the ambient cloud, the geometrically bending structure of the filament, and/or the orientation of the longitudinal magnetic field.
\item The large scale velocity fields of F13 and F13S obtained using the $\tco~(3-2)$ data reveal V-shaped transverse velocity structures, implying F13 and F13s may form through the collision of radiation shock front generated by BD+46 and mass infall by the self-gravity. The mass accretion rates per unit length of F13 and F13S are measured to be $\sim 70 \pm 30~M_{\odot}~\rm pc^{-1}~ Myr^{-1}$ and $\sim 30 \pm 10~M_{\odot}~\rm pc^{-1}~ Myr^{-1}$, and the times for the filaments to become thermally critical are $4 \pm 1 \times 10^{5}$~yr for F13 and $12 \pm 5 \times 10^{5}$~yr for F13S. We propose that the radiation shock might start from BD+46 about 2~Myr ago and collide with the sheet-like molecular cloud $\sim$1.4~Myr ago, and F13 could become critical by the shock compression and self-gravity about 1~Myr ago, and then the dense cores might form by the gravitational fragmentation with the magnetic fields distorted.
\end{enumerate}

\acknowledgments

The authors are grateful to the anonymous referee for useful comments. This research was supported by Basic Science Research Program through the National Research Foundation of Korea (NRF) funded by the Ministry of Education (grant No. NRF- 2022R1I1A1A01053862). C.W.L. was supported by the Basic Science Research Program through the National Research Foundation of Korea (NRF) funded by the Ministry of Education, Science and Technology (NRF- 2019R1A2C1010851), and by the Korea Astronomy and Space Science Institute grant funded by the Korea government (MSIT; project No. 2022-1-840-05). M.T. acknowledges partial support from project PID2019-108765GB-I00 funded by MCIN/AEI/10.13039/501100011033. H.Y. is supported by the Basic Science Research Program through the National Research Foundation of Korea (NRF) funded by the Ministry of Education (NRF-2021R1A6A3A01087238).

%\bibliography{Cocoonpol2ref}

\end{document}

%% file: tbl_fils.tex
\tablecaption{Physical Parameters of the Filaments \label{tbl:fila}}
%\tablecolumns{14}
%\tablewidth{0pt}
\tablehead{
\colhead{} &
\colhead{$T_{\rm dust}$} &
\colhead{$L$} &
\colhead{$W$} &
\colhead{$\bar N_{\rm H_{2}}^{0}$} &
\colhead{$\bar n_{\rm H_{2}}^{0}$} &
\colhead{$M$} &
\colhead{$M_{\rm line}$} &
\colhead{$M_{\rm line}^{\rm th.crit}$} &
%\colhead{$M_{\rm line}^{\rm crit}$} &
\colhead{$\sigma_{\rm NT}$}
\tabularnewline \cline{3-4} \cline{8-9} 
\colhead{} & 
\colhead{(K)} &
\multicolumn{2}{c}{(pc)} &
\colhead{($10^{20}~\rm cm^{-2}$)} &
\colhead{($10^{3}~\rm cm^{-3}$)} &
\colhead{($M_{\odot}$)} &
\multicolumn{2}{c}{($\mspc$)} &
\colhead{($\kms$)}}
\startdata
F13 & 19.8$\pm$1.2 & 1.13$\pm$0.09 & 0.32$\pm$0.07 & 93$\pm$27 & 11$\pm$2 & 76$\pm$28 & 67$\pm$24 & 27$\pm$2 & 0.30$\pm$0.04\\ 
F13S & 24.2$\pm$1.0 & 0.56$\pm$0.05 & 0.23$\pm$0.03 & 42$\pm$12 & 7$\pm$1 & 12$\pm$4 & 22$\pm$7 & 33$\pm$1  & 0.29$\pm$0.05\\ 
\enddata
%\tablecomments{$^{\dagger}$ The physical values are estimated  Filament19 \citep{arzoumanian2011}}

%% file: tbl_alpha.tex
\tablecaption{Fitting Results of the Single Power-law function and the Ricean-mean models \label{tbl:alpha}}
\tablehead{
\colhead{} &
\multicolumn{2}{c}{Single power-law model} & &
\multicolumn{2}{c}{Ricean-mean model} 
\tabularnewline \cline{2-3} \cline{5-6} 
\colhead{} & 
\colhead{$P_{\sigma_{QU}}$} &
\colhead{$\alpha$} & ~~~~~~ & 
\colhead{$P_{\sigma_{QU}}$} &
\colhead{$\alpha$} 
}
\startdata
vectors with $I/\sigma_{I} \geq 10$ \& $P/\sigma_{P} \geq 3$ ~~~~~~~~~~~~~~~~ & 0.64$\pm$0.14 & 0.82$\pm$0.09 &  &  \\ 
\hline
vectors with $I/\sigma_{I} \geq 10$~~~~~~~~~~~~~~~~~~~~~~ &  &  &  & 0.33$\pm$0.08 & 0.72$\pm$0.07\\ 
~~~~~~~~~~close to \bd~$^{\dagger}$ &  &  &  & 0.38$\pm$0.10 & 0.73$\pm$0.07\\ 
~~~~~~~~~~far from \bd~$^{\dagger}$ &  &  &  & 0.29$\pm$0.14 & 0.71$\pm$0.13\\ 
\enddata
\tablecomments{$^{\dagger}$ We divided the vectors into two groups based on their positions within the filaments, one group close to and the other group further away from the B-type star with respect to the skeleton of the filament.}

%% file: tbl_bfield.tex
\tablecaption{Magnetic Field Strengths and Energies \label{tbl:bfields}}
%\tablecolumns{14}
%\tablewidth{0pt}
\tablehead{\colhead{} &
\multicolumn{2}{c}{F13} & &
\multicolumn{2}{c}{F13S}
}
\startdata
$n_{\rm H_{2}}$ ($\rm 10^{3} ~ cm^{-3}$) & \multicolumn{2}{c}{11.0$\pm$2.4} & & \multicolumn{2}{c}{7.0$\pm$0.7} \\ 
$\Delta v$ ($\kms$) & \multicolumn{2}{c}{0.72$\pm$0.08} & & \multicolumn{2}{c}{0.68$\pm$0.11} \\ 
\hline 
& unsharp-masking & structure function & & unsharp-masking & structure function \\ \cline{2-3} \cline{5-6}$\delta \phi$ (degree) & ~~~~9.1$\pm$2.4 & 12.0$\pm$6.2 &  & 16.0$\pm$2.6 & 13.1$\pm$2.1 \\ 
\hline 
$B_{\rm pos}$ ($\mu$G) & ~~~~76$\pm$23 & ~~~~58$\pm$31 &  & 33$\pm$7 & ~~~~40$\pm$9 \\ 
$\lambda$  & ~~~~0.31$\pm$0.13 & 0.40$\pm$0.25 &  & 0.32$\pm$0.12 & 0.26$\pm$0.10 \\ 
$V_{\rm A}$ ($\kms$) & ~~~~1.22$\pm$0.37 & 0.92$\pm$0.50 &  & 0.66$\pm$0.16 & 0.80$\pm$0.19 \\ 
$M_{\rm A}$ & ~~~~0.25$\pm$0.08 & 0.33$\pm$0.18 &  & 0.44$\pm$0.13 & 0.36$\pm$0.10 \\ 
$E_{\rm B}$ (M$_{\odot}$ km$^{2}$ s$^{-2}$) & ~~~~56.4$\pm$26.9 & 32.3$\pm$21.1 &  &2.6$\pm$1.1 & 3.9$\pm$1.6 \\ 
\hline 
$E_{\rm G}$ (M$_{\odot}$ km$^{2}$ s$^{-2}$) & \multicolumn{2}{c}{22.0$\pm$16.1} &  & \multicolumn{2}{c}{1.1$\pm$0.8} \\ 
$E_{\rm K}$ (M$_{\odot}$ km$^{2}$ s$^{-2}$) & \multicolumn{2}{c}{12.3$\pm$4.9} &  & \multicolumn{2}{c}{2.0$\pm$0.8} \\ 
\enddata